\documentclass[acmsmall]{acmart}
\usepackage{wrapfig}
\usepackage{subfig}
\usepackage{algorithm}
\usepackage{multirow}
\usepackage[noend]{algpseudocode}
\AtBeginDocument{%
  \providecommand\BibTeX{{%
    \normalfont B\kern-0.5em{\scshape i\kern-0.25em b}\kern-0.8em\TeX}}}

\setcopyright{acmcopyright}
\copyrightyear{2021}
\acmYear{2021}
\acmDOI{}

\begin{document}

\title{Online Application Guidance for Heterogeneous Memory Systems}
\titlenote{\textbf{Extension of Conference Paper:} While not a true extension, this work builds upon prior contributions published at the IEEE International Conference on Networking, Architecture, and Storage (\textbf{Best Paper Award})~\cite{olson2018nas} and the ACM/IEEE International Symposium on Memory Systems~\cite{olson2019memsys}. The tools and approach developed for these earlier works are described in Section 3. All other sections present original results and contributions.}


\author{M. Ben Olson}
\email{matthew.olson@intel.com}
\affiliation{%
  \institution{Intel\textsuperscript{\textregistered} Corporation}
  \city{Chandler}
  \state{Arizona}
  \country{USA}
}
\author{Brandon Kammerdiener}
\email{bkammerd@vols.utk.edu}
\affiliation{%
  \institution{University of Tennessee}
  \city{Knoxville}
%  \state{Tennessee}
  \country{USA}
}
\author{Kshitij A. Doshi}
\email{kshitij.a.doshi@intel.com}
\affiliation{%
  \institution{Intel\textsuperscript{\textregistered} Corporation}
  \city{Chandler}
  \state{Arizona}
  \country{USA}
}
\author{Terry Jones}
\email{trjones@ornl.gov}
\affiliation{%
  \institution{Oak Ridge National Laboratory}
  \city{Oak Ridge}
  \state{Tennessee}
  \country{USA}
}
\author{Michael R. Jantz}
\email{mrjantz@utk.edu}
\affiliation{%
  \institution{University of Tennessee}
  \city{Knoxville}
%  \state{Tennessee}
  \country{USA}
}

\renewcommand{\shortauthors}{Olson, et al.}

\begin{abstract}
As scaling of conventional memory devices has stalled, many high end and next generation computing systems have begun to incorporate alternative memory technologies to meet performance goals.
Since these technologies present distinct advantages and tradeoffs compared to conventional DDR* SDRAM, such as higher bandwidth with lower capacity or vice versa, they are typically packaged alongside conventional SDRAM in a heterogeneous memory architecture.
To utilize the different types of memory efficiently, new data management strategies are needed to match application usage to the best available memory technology.
However, current proposals for managing heterogeneous memories are limited because they either: 1) do not consider high-level application behavior when assigning data to different types of memory, or 2) require separate program execution (with a representative input) to collect information about how the application uses memory resources. 

This work presents a new data management toolset to address the limitations of existing approaches for managing complex memories.
It extends the application runtime layer with automated monitoring and management routines that assign application data to the best tier of memory based on previous usage, without any need for source code modification or a separate profiling run.
It evaluates this approach on a state-of-the-art server platform with both conventional DDR4 SDRAM and non-volatile Intel\textsuperscript{\textregistered} Optane\textsuperscript{TM} DC memory, using both memory-intensive high performance computing (HPC) applications as well as standard benchmarks. 
Overall, the results show that this approach improves program performance significantly compared to a standard unguided approach across a variety of workloads and system configurations.
The HPC applications exhibit the largest benefits, with speedups ranging from $1.4x$ to $7x$ in the best cases.
Additionally, we show that this approach achieves similar performance as a comparable offline profiling-based approach after a short startup period, without requiring separate program execution or offline analysis steps.
%These extensions enable the runtime layer to: divide the application heap into related sets of application data, collect information about how each set uses memory, and also reassign existing data (and future data through hints for allocation sites) based on its previous usage.
\end{abstract}



\maketitle

%%%%%%%%%%%%%%%%%%%%%%%%%%%%%%%%%%%%%%%%%%%%%%%%%%%%%%%%%%%%
% INTRODUCTION
%%%%%%%%%%%%%%%%%%%%%%%%%%%%%%%%%%%%%%%%%%%%%%%%%%%%%%%%%%%%
\section{Introduction}

Recent years have witnessed the rise of computing systems with more diverse hardware capabilities and heterogeneous architectures.
The memory hierarchy in particular has become more significantly more complex as main memory is now often composed of different device technologies, each with their own performance and capacity characteristics.
For example, Intel\textsuperscript{\textregistered}'s latest Xeon processors (codenamed Cascade Lake) support access to conventional DRAM as well as non-volatile Optane\textsuperscript{TM} DC memory within the same address space.
While this configuration greatly expands the capacity of main memory, access to the non-volatile tier has limited bandwidth and longer latencies.
Other systems, such as the (now defunct) Intel\textsuperscript{\textregistered} Knights Landing (KNL), and many GPU-based platforms, package high bandwidth (but lower capacity) memories (commonly known as ``on-package'' or ``die-stacked'' RAMs) alongside conventional memory to enable better performance for a portion of main memory.
Future memory systems are expected to be even more complex as architectures with three (or more) types of memory and more flexible operating modes have already been announced~\cite{sapphireRapids}.

Despite their potential benefits, heterogeneous memory architectures present new challenges for data management.
Computing systems have traditionally viewed memory as a single homogeneous address space, sometimes divided into different non-uniform memory access (NUMA) domains, but consisting entirely of the same storage medium (i.e., DDR* SDRAM).
To utilize heterogeneous resources efficiently, alternative strategies are needed to match data to the appropriate technology in consideration of hardware capabilities, application usage, and in some cases, NUMA domain.

Spurred by this problem, the architecture and systems communities have proposed a range of hardware and software techniques to manage data efficiently on heterogeneous memory systems.
The existing solutions exhibit various advantages, disadvantages, and tradeoffs, with most hardware-based techniques offering more ease of use and software transparency at the expense of flexibility and efficiency, while software-based solutions provide more fine-grained control of data placement (and thus, better performance) in exchange for additional effort from developers and users.
Section~\ref{sec:related} provides a more detailed overview of these existing approaches.
Unfortunately, there is currently no silver bullet as the more flexible and more efficient software-based approaches still require significant efforts (and, in many cases, expert knowledge) to be effective.

To fill this gap, we began developing a \emph{hybrid} data management solution for complex memory systems based on \emph{automated} application guidance~\cite{olson2018nas,olson2019memsys}.
Our previous approach employs source code analysis and \emph{offline} architectural profiling to collect information about how applications use different regions in their virtual address space.
It also includes a recommendation engine, based on sorting and bin-packing heuristics, to decide which memory tier to use for data allocated during subsequent executions of the same application.
While this approach can significantly improve performance for many applications, it still has some significant limitations.
Specifically, 1) it requires earlier execution with a representative input to collect information about how the application uses program data objects, and 2) it only provides \emph{static} placement recommendations and cannot adjust data-tier assignments as application usage shifts.

This work addresses these limitations by extending our previous approach and toolset with \emph{online} components that are able to collect and apply application-level memory tiering guidance during production execution and \emph{without the need for a separate profile run}.
%We implement this approach as an extension to the high-level interface of the Simplified Interface to Complex Memory (SICM).
%SICM, which is part of the DOE Exascale Computing Project, is a memory allocator and runtime API for adapting HPC applications to heterogeneous memory architectures. 
We evaluate our online approach using HPC as well as standard (SPEC\textsuperscript{\textregistered} CPU) computing benchmarks on an Intel\textsuperscript{\textregistered} Cascade Lake platform with two tiers of memory: conventional DDR4 SDRAM and non-volatile Optane\textsuperscript{TM} DC.
%state-of-the-art heterogeneous memory system with both conventional DRAM and large capacity non-volatile memory. %Intel\textsuperscript{\textregistered} platform with two tiers of memory: conventional DDR4 SDRAM and non-volatile Optane DC memory.
%: 1) collect fine-grained information about the usage rate and capacity requirements of different sets of application data, and 2) steer data placement in consideration of current usage and available resources during production execution and \emph{without the need for a separate profile run}.
Our experiments show that our updated toolset can generate effective tiering guidance with very low overhead and typically achieves performance similar to our previous offline profiling-based approach after a short initial startup period.
The primary contributions of this work are:
\begin{enumerate}
\item{We extend the Simplified Interface to Complex Memory (SICM)\footnote{Described in Section~\ref{sec:sicm}, the SICM project, which is part of the DOE Exascale Computing Project~\cite{exascale-in-us}, is a memory allocator and runtime system designed to facilitate usage of HPC applications on complex memory machines. Its source code adopts and extends the popular jemalloc allocator~\cite{evans2006scalable}.} runtime with new techniques for profiling memory usage during production execution. For the benchmarks in this study, our approach is able to collect detailed data-tiering guidance with negligible execution time overhead in most cases and less than 10\% overhead in the worst case.}
\item{We design and implement an online data tiering solution that leverages this application feedback to steer data allocation and placement across a heterogeneous memory hierarchy. Our approach, inspired by solutions to the classical ski rental problem, only migrates data when the expected cost of doing so is outweighed by the cost of leaving it in place.}
\item{We demonstrate the effectiveness of this approach on a state-of-the-art heterogeneous memory system with both conventional DRAM and large capacity non-volatile RAM. The results show that the online approach significantly outperforms unguided execution on average and achieves speedups ranging from $1.4x$ to more than $7x$ for our selected HPC workloads. We also find that it exhibits similar performance as a comparable offline profiling-based approach after a short initial startup period.}
\end{enumerate}

The remainder of this article is organized as follows.
Section 2 describes related memory management approaches for both heterogeneous and conventional memory machines.
Section 3 presents background information on the tools and software that provide the basis for this work.
Section 4 describes the design and implementation of our approach for adapting this earlier toolset for online execution.
Section 5 describes details of our experimental framework.
Section 6 presents evaluation of the described approach as well as analysis of all results.
Section 7 describes our ongoing and planned efforts to continue and build upon this work, and Section 8 concludes the paper.

%%%%%%%%%%%%%%%%%%%%%%%%%%%%%%%%%%%%%%%%%%%%%%%%%%%%%%%%%%%%
% RELATED WORK
%%%%%%%%%%%%%%%%%%%%%%%%%%%%%%%%%%%%%%%%%%%%%%%%%%%%%%%%%%%%
\section{Related Work}
\label{sec:related}
\subsection{Data Management Strategies for Conventional Systems}
Data placement is a long-standing and well-studied problem in computer science.
Many prior works have successfully used program profiling and analysis to improve data management across the cache, memory, and storage hierarchies.
Some researchers have proposed static techniques with offline profiling and/or source code analysis to allocate hot fields and objects closer together in the heap, thereby improving caching efficiency~\cite{calder1998cache, seidl1998asplos, pool2005alloc, hundt2006cgo, jeon2007cc, hpctoolkit, liu2011asplos}.
Others have combined online profiling with high-level language features, such as object indirection and garbage collection, to enable similar benefits transparently, and in an adaptive runtime environment~\cite{shuf2002oopsla, cherem04region, Huang04locality, guyer2004oopsla, chilimbi2006pldi, hirzel2007sigmetrics, zhang2008ecoop, wang2010cgo, wang2012taco, brock2013ismm, eizenberg2016pldi}.

A number of other works integrate application-level guidance with physical data management in the operating system and hardware.
Some projects developed frameworks or APIs to expose kernel resources to applications~\cite{engler1995exokernel, belay2012dune} or to facilitate communication between upper- and lower-level data management routines~\cite{banga1999containers, kleen2004numa, jantz2013intel, beckmann2013pact}.
More recent efforts have combined these cross-layer approaches with automated collection of high-level guidance to address a variety of issues, including: DRAM energy efficiency~\cite{jantz2015oopsla, olson2018taco}, cache pollution~\cite{guo2015nightwatch}, traffic congestion for non-uniform memories~\cite{dashti2013traffic}, and data movement costs for non-uniform caches~\cite{mukkara2016asplos, tsai2017isca}.
While these works evince some of the benefits of integrating usage feedback during data management, their purposes and goals are very different from this project.

\subsection{Data Management Strategies for Heterogeneous Memory Systems}
Propelled by the simultaneous growth of data analytics and stalling of conventional DRAM scaling, research interest in alternative memory technologies has grown significantly in the last decade.
The shifting landscape has pushed the architecture, systems, and high-performance computing communities to propose new strategies, tools, and techniques for mapping application data across heterogeneous device tiers.
\vspace{-1ex}
\subsubsection{Hardware-Managed DRAM Caches:}
One common strategy is to exercise the faster, smaller capacity tier(s) as a hardware-managed cache.
For example, Intel\textsuperscript{\textregistered}'s Cascade Lake includes a ``memory-mode'' option, which applies this approach with DDR4 as a direct-mapped cache to storage class Optane\textsuperscript{TM} DC  memory~\cite{optane2019swanson}.
While hardware-managed caching provides some immediate advantages, such as software-transparency and backwards compatibility, it is inflexible, often less efficient, and reduces the system's available capacity.
%It also imposes unpalatable architectural costs as the high performance tier must either be implemented as a tagless direct mapped (non-associative) cache, or it requires logic and storage for associative tags.
%These issues become even more problematic as the capacity of the hardware-managed tier(s) increases, and so scalability of this technique is a concern.

Some works have proposed architectural strategies to address these issues, for example, by: co-locating tags and data in DRAM to increase efficiency~\cite{loh2011efficiently, meza2012enabling}, keeping track of cache contents in TLBs and page tables to reduce metadata traffic~\cite{lee2015isca, jang2016hpca, young2017isca}, or swapping data lines out of the cache to preserve capacity~\cite{chou2014micro, sim2014micro}.
%been exploring the design challenges and tradeoffs of architecting DRAM as a
%large capacity last level cache
Mittal and Vetter provide a modern (2016) survey of this research~\cite{mittal2016survey}.
In contrast to these works, this work extends and develops techniques to increase efficiency solely through software-driven data placement, without relying on architectural modifications or non-standard hardware.
Some recent work has also shown that profile guidance can enhance data management on systems which support hardware-directed caching and OS paging \emph{simultaneously}, but for different portions of their address space~\cite{effler2020mchpc}.
We expect the approach proposed in this work can also boost performance on platforms with such mixed data management options.

\paragraph{Software-Directed Heterogeneous Memory Management:}
The alternative strategy of \emph{software-based} data tiering uses either the OS by itself, or the OS in conjunction with the application to assign data into different memory tiers, with facilities to allow migrations of data between tiers as needed.
%Implementations of this approach are often similar to data management on NUMA
%platforms~\cite{kleen2004numa, mbind}, with each tier represented as its own
%NUMA domain.
Some heterogeneous memory systems also provide API's that allow applications to control the placement of their data objects through the use of source code annotations~\cite{memkind, pascal}.
These finer-grained controls enable developers to coordinate tier assignments with data allocation and usage patterns, potentially exposing powerful efficiencies.
%However, software-based approaches require expert knowledge and are often less adaptive due to the need to synchronize low-level data movement with the application.
%but require expert knowledge and
%manual modifications to source code.
%However, while virtualization provides some flexibility, software-based
%approaches are often less adaptive due to the need to synchronize low-level
%data movement with the application.
%with facilities to allow migration of data
%between those tiers as needed.

Several prior works have integrated software-based data management with program profiling to facilitate the assignment of data to memory tiers.
For instance, some prior works integrate coarse-grained architectural profiling with page-level management in the OS~\cite{Meswani15dieStack, mutlu2017cluster, agarwal2017asplos, kim2021atc}.
Since these works do not attempt to coordinate tier assignments with application data structures and events, they may be vulnerable to inefficiencies that arise from the high-level software working at cross-purposes from the OS and hardware.
%In contrast, our approach will coordinate hardware monitoring with static
%analysis and lightweight runtime support to enable online and fully automatic
%data-tiering guidance from the application.

Some other projects employ application-level tools to tag and profile certain data structures, and then use heuristic models to assign objects to the appropriate tier~\cite{agarwal2015asplos, dulloor2016data, peng2017ismm, servat2017cluster, laghari2018phase, akram2021taco}.
While these efforts demonstrate that application guidance can be useful for certain usage scenarios, they require manual source code modifications or expensive online detection to attach recommendations to data objects.
Several prior works, including our own, have attempted to address this limitation with static and lightweight runtime tools that are able to attach tiering guidance to program data \emph{automatically}~\cite{unimem2017sc, effler2018arcs, olson2018nas, olson2019memsys, effler2020mchpc}.
%To address this limitation, our preliminary work has several static and lightweight runtime tools to attach guidance to program data \emph{automatically}~\cite{olson2018taco, effler2018arcs, olson2018nas, valence}
%developed static and lightweight runtime tools to attach guidance
%to program data \emph{automatically}.
However, all of these previous works employ \emph{offline} profiling and analysis to collect information about how the application uses memory, and generate only static tier recommendations.
%Some studies~\cite{unimem2017sc, laghari2018phase} have
%implemented adaptive placement policies, but these approaches still
%require offline profiling and migrate only a small number of data
%objects at relatively infrequent intervals to manage overheads.
In contrast, this project leverages lightweight architectural profiling and novel runtime algorithms to enable automated, feedback-directed data placement with very low execution time overhead.
Moreover, it does so without requiring earlier, profiled execution of the same application with representative input.

\section{Offline Application Guidance for Heterogeneous Memory Systems}
The \emph{online} data tiering approach described in this work builds upon our earlier efforts to improve application performance on heterogeneous memory systems.
Our previous work extended the SICM runtime and API to implement an \emph{offline} profile-based approach for guiding data placement on multi-level memory systems.
This section provides a brief overview of SICM as well as our offline approach, which is called MemBrain.
%We recently combined our high-level offline approach, which is called MemBrain, with the SICM runtime and API to create a portable application guidance framework, which forms the basis of this work.
%This section provides a brief overview of the primary components of our offline framework: our high-level approach, which is called MemBrain, and the SICM runtime system and API, which provides the basis of our implementation, and .

\begin{wrapfigure}[17]{r}{0.5\textwidth}
  \vspace{-2ex}
  \centering
  \includegraphics[width=\linewidth]{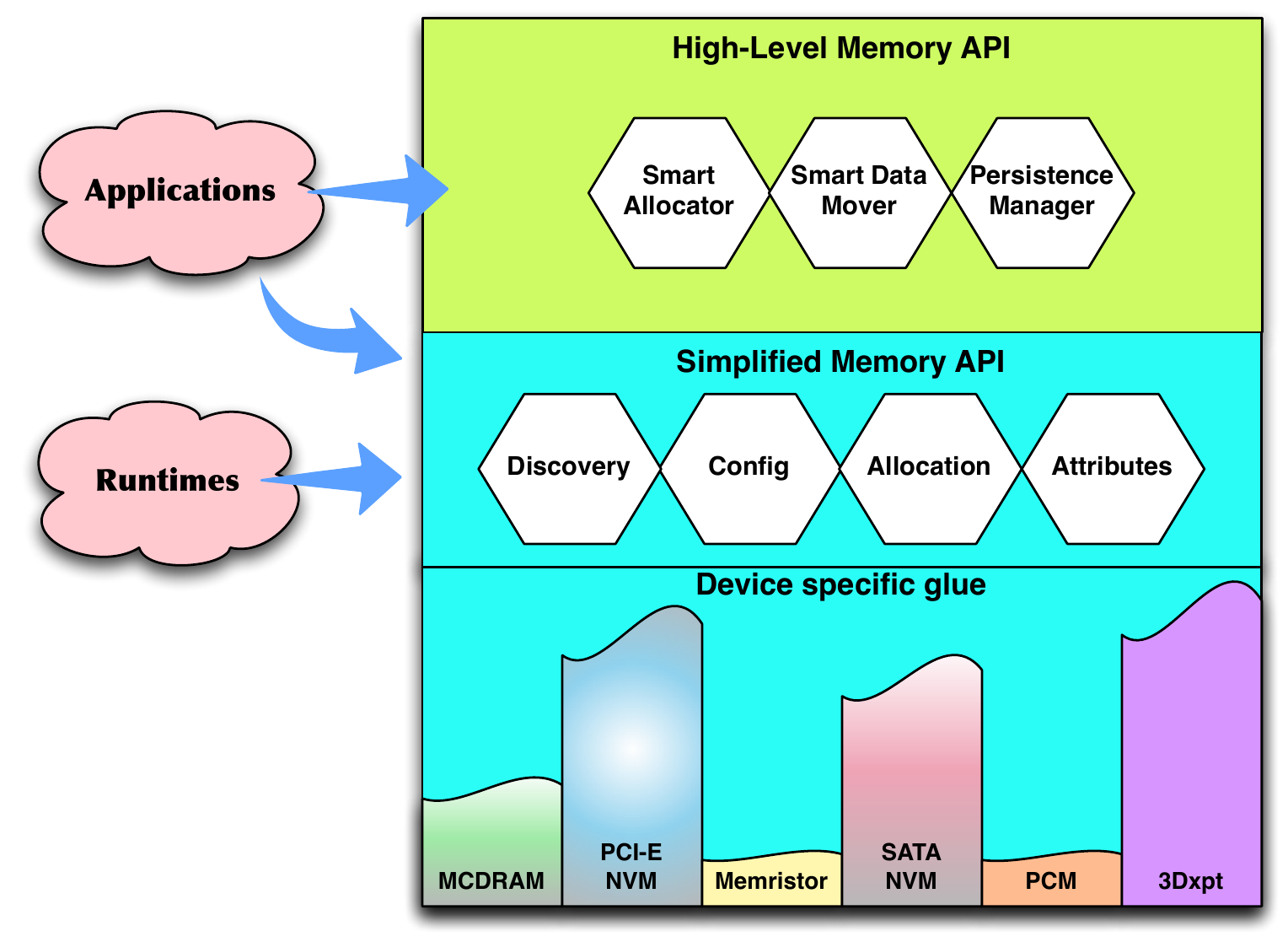}
  \captionof{figure}{SICM overview \cite{sicm}.
  The high-level provides a portable API, while the
  low-level implements efficient data management for
  complex memories.}
  \label{fig:sicm}
\end{wrapfigure}
\vspace{-1ex}
\subsection{Simplified Interface to Complex Memory}
\label{sec:sicm}
%The U.S. Department of Energy (DOE) is working towards achieving new levels of scientific discovery through ever-increasingly powerful supercomputers~\cite{aurora,frontier}.
%Short-term plans call for achieving exaFLOP performance by the year 2021.
The U.S. Department of Energy (DOE) Exascale Computing Project (ECP) is a large, multi-disciplinary effort with the goal of achieving exaFLOP performance in the supercomputing domain~\cite{exec-order,exascale-in-us}.
%The project includes multiple thrust areas to deal with the hardware and software challenges of the most complex and high-performance supercomputers.
%For such systems, the DOE spends hundreds of millions of dollars to achieve the highest performance possible from available hardware.
The Simplified Interface to Complex Memory (SICM) is is one of the ECP subprojects.
It seeks to deliver a simple and unified interface to the emerging complex memory hierarchies on exascale nodes~\cite{sicm,aurora,frontier}.
To achieve this goal, SICM employs two separate interfaces, as shown in, as shown in Figure~\ref{fig:sicm}.
The high-level interface delivers an API that allows applications to allocate, migrate, and persist their data without detailed knowledge of the underlying memory hardware.
To implement these operations efficiently, the high-level API invokes the low-level interface, which interacts directly with device-specific services in the OS.
Our prior work extends both layers of SICM with profiling tools and analysis, as well as new data management algorithms, to enable guided data placement on complex memory platforms~\cite{olson2019memsys}.
%The following new work is implemented in SICM's framework,
%making use of its low-level interface and extending its 
%high-level interface to enable profile-guided allocations.

\vspace{-1ex}
\subsection{MemBrain: Automated Application Guidance for Hybrid Memory Systems}
\begin{figure}[t]
  \centering
  \includegraphics[width=\linewidth]{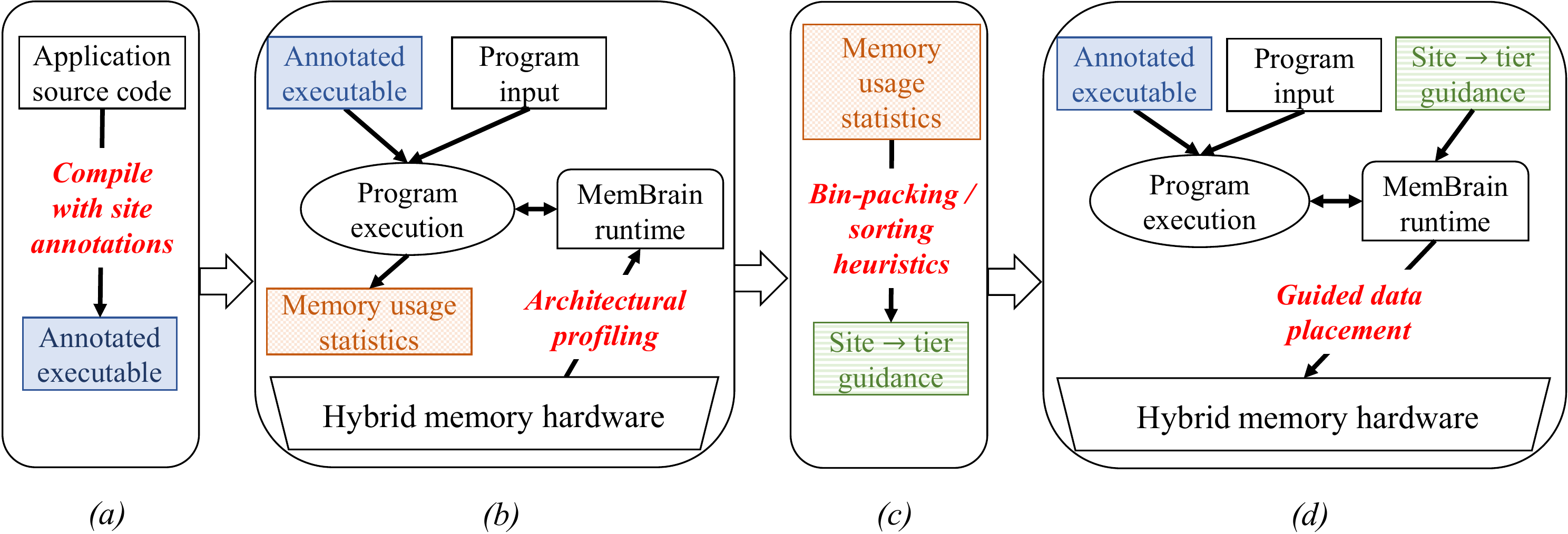}
  \captionof{figure}{Data tiering with offline application
  guidance~\cite{olson2018nas}. (a) Compile executable with
  source code annotations at each allocation site, (b)
  Profile memory usage of each site in a separate
  program run using architectural sampling, (c) Employ
  bin-packing / sorting heuristics to assign data-tier
  recommendations to each site, (d) Apply data-tiering
  recommendations during subsequent program executions.}
  \label{fig:membrain}
\end{figure}

To automate the conversion of program profiles to tier recommendations for different memory regions, this work adopts a similar strategy as our previous offline approach called MemBrain~\cite{olson2018nas}.
MemBrain generates data-tier guidance by associating profiles of memory behavior (such as bandwidth and capacity) with program \emph{allocation sites}.
Each allocation site corresponds to the source code file name and line number of an instruction that allocates program data (e.g., \texttt{malloc} or \texttt{new}) and may optionally include part or all of the call path leading up to the instruction.
A separate analysis pass converts the profiles into tier recommendations for each site prior to guided execution.
Figure~\ref{fig:membrain} presents an overview of this approach.

\vspace{-1ex}
\subsubsection{Converting Site Profiles to Tier
Recommendations}
\label{sec:binpack}
MemBrain includes three options for converting memory usage profiles into tier recommendations for each allocation site:
%The three options are as follows:
%For this work, we implement and evaluate all three of
%the following options in our portable application
%guidance framework:
\vspace{-1ex}
\paragraph{Knapsack:} The knapsack approach views the task of assigning application data into different device tiers as an instance of the classical 0/1 knapsack optimization problem.
In this formulation, each allocation site is an item with a certain value (bandwidth) and weight (capacity).
The goal is to fill a knapsack such that the total capacity of the items does not exceed some threshold (chosen as the size of the upper tier), while also maximizing the aggregate bandwidth of the selected items.

\vspace{-1ex}
\paragraph{Hotset:} The hotset approach aims to avoid a weakness of knapsack, namely, that it may exclude a site on the basis of its capacity alone, even when that site exhibits high bandwidth.
Hotset simply sorts sites by their bandwidth per unit capacity, and selects sites until their aggregate size exceeds a soft capacity limit.
For example, if the capacity of the upper tier is C, then hotset stops adding the sorted sites after the total weight is just past C.
By comparison, knapsack will select allocation sites to maximize their aggregate value within a weight upper bound of C.
%knapsack will select allocation sites so that their
%aggregate size is just below C, while hotset stops adding
%sites after the total size is just past C.

\vspace{-1ex}
\paragraph{Thermos:} Since hotset (intentionally) over-prescribes capacity in the upper tier, cold or lukewarm data could potentially end up crowding out hotter objects during execution.
The thermos approach aims to address this occasional drawback.
It only assigns a site to the upper tier if the bandwidth (value) the site contributes is greater than the aggregate value of the hottest site(s) it may displace.
In this way, thermos avoids crowding out performance-critical data, while still allowing large-capacity, high-bandwidth sites to place a portion of their data in the upper-level memory.

\section{Online Application Guidance for Heterogeneous Memory Systems}
\label{sec:online_profiling}
Our earlier approaches for guiding data tiering are limited because they require a separate, profiled execution of each application (with representative input) and only generate static data-tier recommendations for subsequent program runs.
This work addresses these limitations by adapting MemBrain for use as an online and fully automated feedback-directed optimization.
Specifically, our updated approach monitors application memory behavior, converts this information into data-tier recommendations, and enforces these recommendations to distribute data efficiently across the memory hierarchy, all within a single run of the application.

Realizing this vision required two major extensions to our existing SICM+MemBrain framework:
1) Updates to the profiling infrastructure, including new arena allocation schemes and OS instrumentation, to increase the efficiency of collecting and organizing memory usage information, and
2) A new online decision engine that analyzes the profiles of all active memory regions and decides when and how to migrate application data across the available memory hardware.
This section presents design and implementation details for these new components.

\subsection{Effective Memory Usage Monitoring with Low Overhead}
The earlier MemBrain approach attempts to provide memory tier recommendations for the data associated with each program allocation context.
To do so, it requires two bits of information for each allocation context: 1) the cumulative resident set size (RSS) of the data it allocates, and 2) the usage rate of its data relative to other contexts.
%To collect this information, Our earlier works employed an offline profiling run to collect this information for different sets of application data, where each set corresponds to the all of the data allocated from the same allocation context. 
To collect this information, it employs an offline profile run where each allocation context is associated with a distinct page-aligned region of virtual addresses, known collectively as an \emph{arena}.
During the profile run, each new data object is allocated to an arena that is unique to its own allocation context.
This approach ensures objects from different allocation contexts do not share the same page, which facilitates profile collection.

To estimate the relative access rate of the data in each arena, our profiling tools employ architectural features commonly available in modern processors.
Specifically, the profiler uses the Linux perf facility~\cite{perf} to sample the addresses of data accesses that miss the last level cache (LLC).
%Specifically, the profiler uses the Linux perf facility~\cite{perf} to sample the addresses of data accesses that miss the last level cache (LLC).
It then maps each sampled address to its corresponding arena and maintains a count of the number of accesses to data in each arena.
In this way, the counts comprise a heatmap of the relative usage of each allocation context at the end of the profile run.
Additionally, the profiler estimates the maximum resident set size of each allocation context by keeping track of the number of physical pages associated with each arena.
For this work, we have updated this mechanism to reduce its overhead, as described in Section~\ref{sec:rss}.

\begin{figure}[t]
  \centering
  \includegraphics[width=\linewidth]{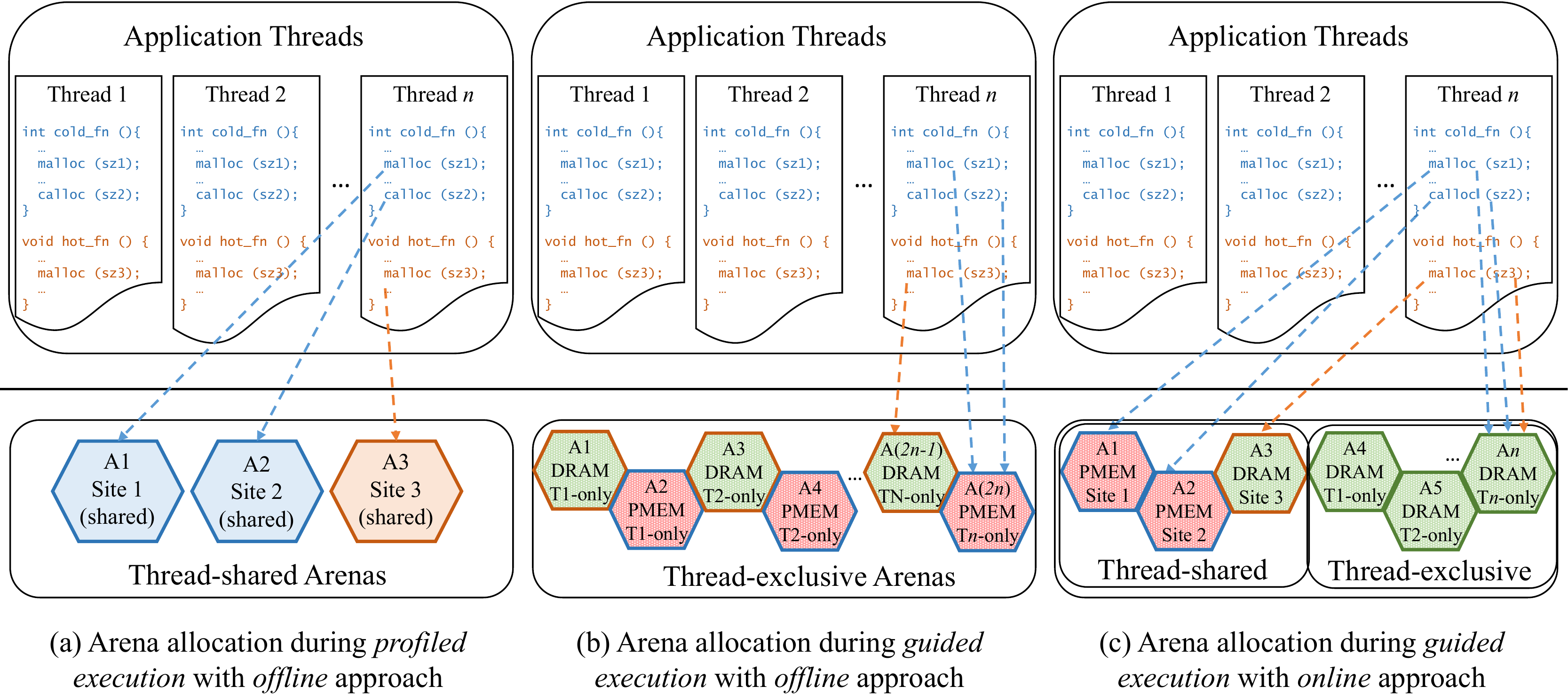}
  \captionof{figure}{Arena allocation strategies for the \emph{offline} and \emph{online} data tiering approaches. The dashed and dotted lines show how the \emph{n\textsuperscript{th}} program thread allocates data from each allocation instruction. In (b), the first and second allocation instructions always use an arena backed by physical memory in the faster DRAM tier because prior profiling indicates the data created at these sites are accessed frequently. In (c), each allocation instruction will use a thread-exclusive arena until the total bytes allocated by the instruction exceeds a predefined threshold. After this point, the instruction will use a shared arena, which may be remapped to different memory tiers over time, depending on the current profile and tier recommendations.}
  \label{fig:arenas}
\end{figure}
\subsubsection{Hybrid Arena Allocation to Reduce Locking}
\label{sec:hybrid_arenas}
While our earlier approach is suitable for offline profiling, it can incur significant execution time overheads (more than $2x$, in some cases) and often takes too long to build effective guidance for usage in an online feedback-directed optimization (FDO).
On further investigation, we found that most of the execution time overhead is due to thread contention during allocation to shared arena spaces.
In our original profiling configuration, all application threads that allocate from the same program context use the same arena, as shown in Figure~\ref{fig:arenas}(a).
If two or more threads try to allocate from the same context simultaneously, one thread will acquire a lock and force the other threads to wait while it completes its allocation request.
While such locking can degrade the performance of the profile run, the slowdowns can be avoided during subsequent guided executions by using a different arena allocation strategy. 
In our original offline approach, the guided run creates a unique set of arenas for every program thread (i.e., one arena for each memory hardware tier) to obviate the need for locking, as shown in Figure~\ref{fig:arenas}(b).

However, this strategy is not feasible for an entirely online approach where profiling is performed alongside guided execution.
Moreover, the na\"{i}ve approach of creating a unique set of arenas for every allocation context for every thread is also not sufficient because many of the applications that run on complex memory hierarchies employ dozens of threads and reach hundreds or thousands of allocation contexts.
Hence, creating thousands of unique arenas for every program thread fragments the address space and reduces spatial locality.
It also slows down operations to aggregate and analyze memory usage profiles of each arena, which can also reduce the efficacy of this approach.

To address these issues, we developed a \emph{hybrid} arena allocation scheme that aims to enable profiling of most application data without the need for locking in most cases.
Our approach exploits the observations that: 1) most of the lock contention during profiling arises due to frequent allocations of very small data objects, and 2) even if they are cold or their usage patterns unknown, such allocations can often be assigned to the smaller, faster tier(s) of memory with little penalty since they do not require much capacity.

Figure~\ref{fig:arenas}(c) presents our hybrid allocation scheme.
The allocator for the hybrid scheme creates two sets of arenas: one set of thread private arenas, each of which may contain data created from any allocation context, and another set of arenas shared among all program threads, each of which corresponds to exactly one allocation context.
By default, all program data is allocated to the private arena corresponding to the thread that created it.
However, the runtime also keeps track of the cumulative size of the data allocated at each allocation context.
When the number of active bytes corresponding to a particular context exceeds a predefined threshold (say, 4 MB), new data created from that context are allocated to the shared arena designated for that context.

In this way, frequent allocations from contexts with smaller capacity requirements can complete without needing to lock a shared resource.
Additionally, by choosing an appropriately small threshold, the private arenas will never require much physical capacity, and can always be assigned to the smaller, faster tier(s) with little penalty.
Hence, the online profiler does not attempt to track the origin of data in the thread private arenas, and only profiles the usage of the shared arenas.

\subsubsection{System-Level Integration for More Effective Capacity Profiling}
\label{sec:rss}
Another challenge in adapting the SICM+MemBrain approach for use as an online FDO is that the approach it uses to measure the capacity requirements of each arena can incur significant overheads and is often too slow to be effective.
Specifically, our previous approach employed a separate runtime thread to periodically count up the number of resident physical pages using the Linux pagemap facility~\cite{pagemap}.
There are two main drawbacks to using this approach in an online framework: 1) to prevent the application from modifying addresses as they are read, the profiling thread has to lock each arena as it walks over the heap, and 2) it can be very slow for large applications because it requires numerous seek and read system calls to collect information about each and every virtual page.

For this work, we developed an alternative strategy that leverages existing data structures and deeper integration with the Linux kernel to enable fast and effective capacity profiling for large scale applications.
Linux organizes the virtual address space of each process into a set of Virtual Memory Areas (VMAs), where each VMA is comprised of a contiguous range of virtual addresses with similar access permissions and other properties.
The metadata for each region is kept in a structure called the \texttt{vm\_area\_struct}, and information regarding each VMA, such as its address range, backing file, and permissions, can be read by applications via the \texttt{\/proc} interface.

For this enhancement, we extended Linux's \texttt{\/proc} interface with facilities for applications to create a new VMA for a given virtual address range, provided that the given range is already part of the process's virtual address space.
% MBO: This section mentions that OBJMAP includes a custom system call, but there is no new system call:
% it's just a `proc` interface that supports writing/reading the files in it; when you write a new file,
% a range of addresses becomes a new VMA, and when you read from that file (whose name corresponds to the
% start and end addresses), the kernel modification executes some code which determines how many
% resident pages there are in that range and returns the value back to you.
Additionally, we added instrumentation in the page fault and page release paths of the Linux memory manager to track the number of resident physical pages corresponding to each VMA.\footnote{This instrumentation is actually straightforward to implement in recent Linux kernels as it follows existing code to track of the resident set size of each memory control group.}
To track the RSS of each arena, the application runtime creates new VMAs for each contiguous range of addresses within each arena  by writing to the custom \texttt{\/proc} interface.
The online profiling thread then reads from this same \texttt{\/proc} interface to collect up-to-date counts of the number of resident pages for each VMA (and by extension, each arena).

While our current implementation relies on Linux kernel modifications, some recent features make it possible to implement this approach without any changes to kernel code.
For example, the extended Berkeley Packet Filter (eBPF), which has been supported in Linux since version 4.1, enables users to write and attach custom instrumentation to a live kernel image, without any risk of crashing or hanging system code~\cite{ebpf}.
In the future, we plan to remove dependence on kernel modifications by adapting our approach to use eBPF to track resident memory for each arena.

\vspace{-1ex}
\subsection{Deciding When and How to Migrate Application Data}
An important component of any online FDO is how it decides if and when to expend computing resources on program optimization.
Optimizing too early can lead to poor optimization decisions due to inaccurate or incomplete profile information.
Such premature optimization is especially harmful in the context of this work due to the high cost of migrating data across memory tier boundaries.
However, optimizing too late is also harmful because the program will spend a longer portion of its execution time without the benefit of the optimization. 
Previous works that use offline profiling or static tier recommendations avoid this dilemma because the information needed to optimize is readily available at the start of program execution.

To construct an online FDO for data tiering, we express the problem of choosing when to migrate application data as an instance of the classical ski rental problem.
The ski rental problem describes a class of optimization problems where, at every time step, one must pay a repeating cost (i.e., renting a pair of skis) or pay a larger one-time cost to reduce or eliminate the repeating cost (i.e., buying a pair of skis).
This formulation has been used to solve online problems in a range of domains including just-in-time (JIT) compilation~\cite{brock2018cc}, cache coherence~\cite{karlin1990soda}, and cloud computing~\cite{khanafer2013infocom}.
For this work, we view the problem of whether to move application data across tiers as a choice between continuing to pay the repeating cost of keeping relatively warm data in a slow memory tier and paying the larger cost of remapping application data to a different tier.

\begin{algorithm}
\caption{Online Guided Data Tiering. The constants $EXTRA\_NS\_PER\_SLOWER\_ACCESS$ and $NS\_PER\_PAGE\_MOVED$ are roughly equal to the average additional latency per data access on the slower memory tier (in ns), and the average execution time cost (in ns) of remapping a single virtual page from one tier of memory to the other, respectively.}
\begin{algorithmic}[1]
\Procedure{GetRentalCost}{$prof$, $recs$}
  \State $rentalCost \gets a \gets b \gets 0$;
  \For {($site$, $curTier$, $accs$, $pages$) \textbf{in} $prof$}
    \State $recTier \gets$ GetRecTier($site$, $recs$);
    \If {$curTier = OPTANE\_TIER$ \textbf{and} $recTier = DRAM\_TIER$}
      \State $a \gets a + accs$;
    \ElsIf {$curTier = DRAM\_TIER$ \textbf{and} $recTier = OPTANE\_TIER$}
      \State $b \gets b + accs$;
    \EndIf
  \EndFor
  \If {$a > b$}
    \State $rentalCost \gets ((a-b) * EXTRA\_NS\_PER\_SLOWER\_ACCESS)$;
  \EndIf
  \State \textbf{return} $rentalCost$;
\EndProcedure
\State
\Procedure{GetPurchaseCost}{$prof$, $recs$}
  \State $totalPagesToMove \gets 0$;
  \For {($site$, $curTier$, $accs$, $pages$) \textbf{in} $prof$}
    \State $recTier \gets$ GetRecTier($site$, $recs$);
    \If {$curTier = DRAM\_TIER$ \textbf{and} $recTier = OPTANE\_TIER$}
      \State $totalPagesToMove \gets totalPagesToMove + pages$;
    \ElsIf {$curTier = OPTANE\_TIER$ \textbf{and} $recTier = DRAM\_TIER$}
      \State $totalPagesToMove \gets totalPagesToMove + pages$;
    \EndIf
  \EndFor
  \State \textbf{return} $(totalPagesToMove * NS\_PER\_PAGE\_MOVED)$;
\EndProcedure
\State
\Procedure {MaybeMigrate}{}
  \State $prof \gets$ CollectCurrentProfile ();
  \State $recs \gets$ GetTierRecs ($prof$); \Comment {uses one of the MemBrain approaches (Sec.~\ref{sec:binpack})}
  \State $rentalCost \gets$ GetRentalCost ($prof$, $recs$);
  \State $purchaseCost \gets$ GetPurchaseCost ($prof$, $recs$);
  \If {$rentalCost > purchaseCost$}
    \State EnforceTierRecs($recs$); \Comment{remap arenas (invokes \texttt{move\_pages} via SICM)}
  \EndIf
\EndProcedure
\State
\Procedure{OnlineGDT}{} \Comment {entry point, starts in a separate runtime thread}
  \State EnableProfiling ();
  \While {True}
    \State Wait ($IntervalTime$);
    \State MaybeMigrate ();
    \State ReweightProfile (); \Comment{optionally reweight to ``forget'' older profile information}
  \EndWhile
\EndProcedure
\end{algorithmic}
\label{alg:online}
\end{algorithm}
Our solution follows the break-even algorithm, which is known to be the best deterministic algorithm for solving the ski rental problem~\cite{skirental}.
Algorithm~\ref{alg:online} presents pseudocode of our approach.
As the application executes, a separate runtime thread counts the total number of memory access samples and number of pages resident on each memory tier in each virtual arena.
The runtime then examines this information at regular intervals to determine if and how it should move any data to a different memory tier.
For this operation, it first estimates the \emph{optimal} data-tier assignments for every arena and allocation site using one of the three MemBrain strategies (i.e., knapsack, hotset, or thermos) with the current memory profile.
Next, it computes and compares two costs: 1) the rental cost, which is the expected cost of keeping the current data-tier assignments, and 2) the purchase cost, which is the cost of migrating application data to match the MemBrain recommendations.

To compute the rental cost, our approach calculates $(a)$ the number of data reads that are resolved in the slower memory tier, but which would have been resolved on the faster memory if the optimal data-tier assignments were enforced, as well as $(b)$ the number of reads resolved in faster memory that would have been resolved in slower memory with the recommended data placement.
The runtime can estimate these values online by scaling the relevant sample counts in the current profile by the sample period.
If $(a)$ is greater than $(b)$, then $(a - b)$ is also multiplied by the average additional latency necessary to read data from the slower devices.
For example, on our experimental platform, the average read latency of the Optane\textsuperscript{TM} DC tier is about 300ns longer than the DDR4 SDRAM memory tier~\cite{optane2019swanson}.
Thus, the rental cost is calculated as $(a - b) * 300ns$.

\begin{figure}[t]
  \centering
  \includegraphics[width=0.65\linewidth]{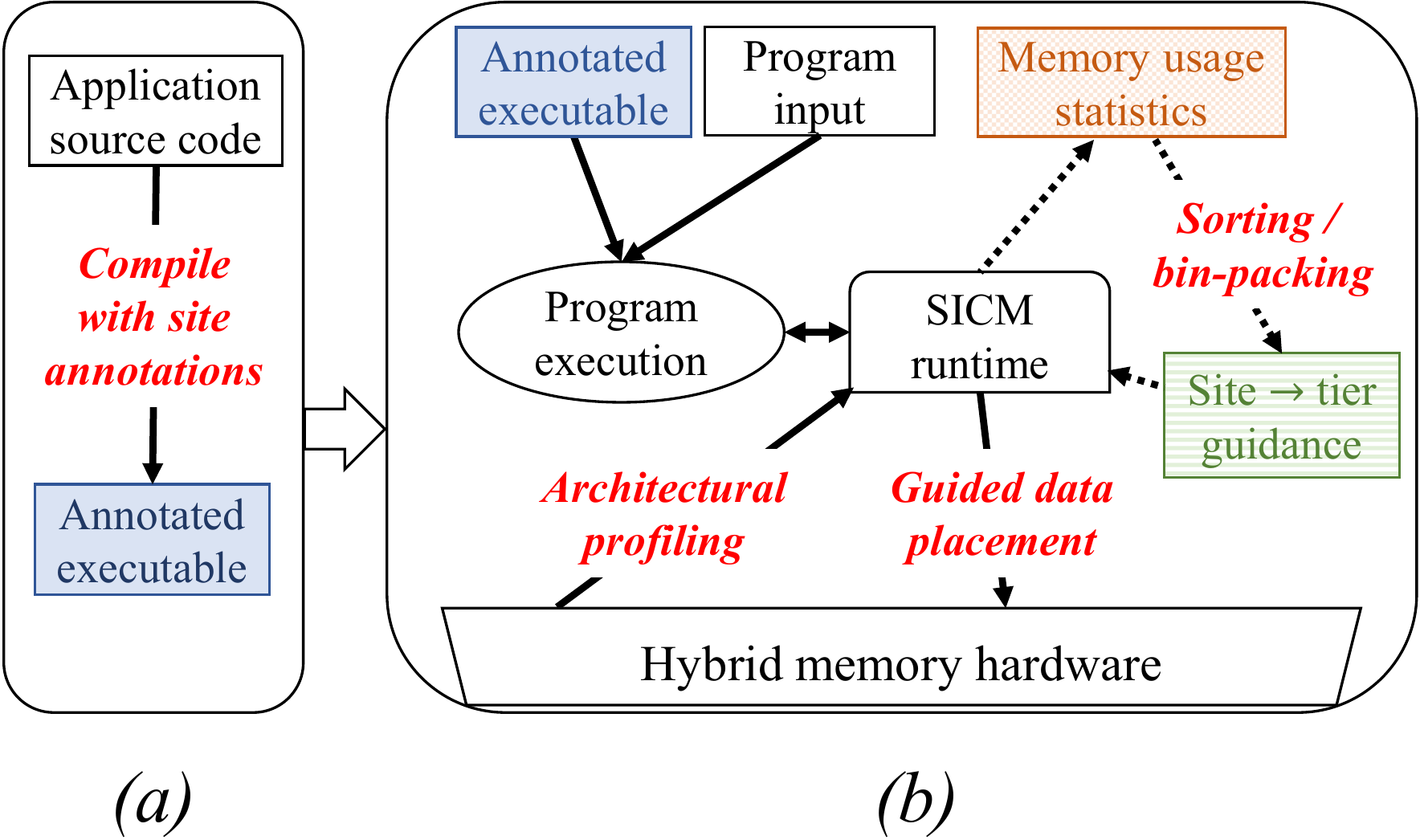}
  \captionof{figure}{Data tiering with online application
  guidance. (a) Users first compile the application with a custom pass to insert annotations at each allocation call site, (b) Program execution proceeds inside a custom runtime layer, which automatically profiles memory usage behavior, converts it into tier recommendations for each allocation site, and enforces these recommendations during program execution. In (b), interactions and operations drawn with dashed lines only occur at regular, timer-based intervals, while the solid lines correspond to activities that can occur throughout the program execution.}
  \label{fig:online_overview}
\end{figure}

To estimate the purchase cost, the runtime computes the number of pages of data it would need to move to enforce the optimal tier recommendations, and multiplies this value by the average rate with which the platform is able to migrate data between tiers.
On our Linux-based platform, we found that moving data between tiers (via the \texttt{move\_pages} system call) requires about 2$\mu$s for each 4 KB page.
Thus, we estimate the purchase cost as $2\mu$s times the total number of pages that would move if the tier recommendations were to be enforced.

At each decision interval, the runtime simply compares the rental and purchase costs.
If the cumulative rental cost ever exceeds the purchase cost, the current data-tier recommendations are enforced.
%relatively cold data that would need to move from the faster tier to the slower tier to make space, as well as relatively warm data that would need to move from the slower tier to the faster tier.
Specifically, any arenas that are mapped to the faster tier and that contain relatively cold program data will first be remapped to the slower tier to make space for the hotter data.
Next, arenas with relatively warm program data residing in the slower tier will then be remapped to the faster tier.
Additionally, the runtime updates a side table with the current site-tier assignments to ensure accurate bookkeeping going forward.

Before completing the interval, the runtime may optionally reset or re-weight the profile information to gradually forget older intervals and enable faster adaptation to new memory behavior.
However, in our current implementation, profile information is never reset or re-weighted.
Memory access samples always accumulate over time, and capacity estimates are updated instantaneously as the application maps and unmaps physical memory in its address space.
We have found that this configuration tends to work well for applications with relatively stable memory usage patterns, including most of the memory-intensive applications we used for this study.

\vspace{-1ex}
\subsection{Summary of Online Approach}
Figure~\ref{fig:online_overview} shows an overview of our online approach.
The online approach still employs compiler analysis to annotate each allocation instruction, and potentially several layers of function call context, with a unique identifier.
Once the annotated executable has been built, the profile-guided data tiering process is entirely automatic.
During program execution, the custom runtime collects memory usage information, converts it to tier recommendations for the application's allocation sites and existing program data, and enforces these data-tier recommendations, all within the same program run, and without any need for additional input or direction from the user.

\section{Experimental Setup}
\subsection{Platform Details}
Our evaluation platform contains a single Intel\textsuperscript{\textregistered} Xeon\textsuperscript{\textregistered} Gold 6246R processor (codenamed ``Cascade Lake'' or CLX)
with 16 compute cores, each running with a clock speed of 3.4 GHz, and a shared 35.75 MB L3 cache.
Its memory system includes 192 GB (6 $x$ 32 GB) of 2933 MT/s DDR4 SDRAM and 768 GB (6 $x$ 128 GB) of 2666 MT/s Optane\textsuperscript{TM} DC persistent memory.
For data reads, the Optane\textsuperscript{TM} tier requires $2x$ to $3x$ longer latencies and sustains 30\% to 40\% of the bandwidth as the DDR4 memory.
While latency for writes is similar on both tiers, the DDR4 tier supports $5x$ to $10x$ more write bandwidth than the Optane\textsuperscript{TM} tier~\cite{optane2019swanson}.

We installed Debian 10 with Linux kernel version 5.7.2 as the base operating system.
For all software-based tiering configurations (i.e., \emph{first touch}, \emph{offline}, and \emph{online}), we used system configuration tools (e.g., \texttt{daxctl}, \texttt{ndctl}, etc.), to assign the DDR4 and Optane\textsuperscript{TM} device tiers to separate NUMA nodes.
This configuration allows applications and system software to track and modify allocations to each type of memory using the standard NUMA API~\cite{kleen2004numa}.

\begin{table*}[t]
  \tiny
  \caption{Workload descriptions and statistics. The columns on the right show the \# of allocation sites reached during execution, as well as the execution time and peak resident set size of each benchmark with the default (unguided first touch) configuration. The CORAL table also shows the arguments that were used to construct the different inputs for each workload as well as the absolute figure of merit (FoM) for the default configuration. All experiments with the SPEC\textsuperscript{\textregistered} CPU 2017 benchmarks use the standard \emph{ref} program input.}
  \centering
  \begin{tabular}{|c|p{3.6cm}|c|l|r|r|r|r|}
    \hline
    \multicolumn{8}{|c|}{CORAL}\\
    \hline
    \multicolumn{1}{|c|}{\textbf{Application}}       &
    \multicolumn{1}{c|}{\textbf{Description}}        &
    \multicolumn{1}{c|}{\textbf{Input}}              &
    \multicolumn{1}{c|}{\textbf{Input Arguments}}    &
    \multicolumn{1}{c|}{\textbf{FoM}}                &
    \multicolumn{1}{c|}{\textbf{Time}}               &
    \multicolumn{1}{c|}{\textbf{GB}}                 &
    \multicolumn{1}{c|}{\textbf{Sites}}              \\
    \hline
    
    \multirow{3}{*}{LULESH} &
    \multirow{3}{*}{\parbox{3.6cm}{Hydrodynamics
    stencil calculation, very little communication between
    computational units. \textbf{FoM: zones per second}}}
    & Medium & \texttt{-s 400 -i 6 -r 11 -b 0 -c 64 -p} & 1,066.93 & 6.2m & 66.2 & 87 \\
    & & Large & \texttt{-s 800 -i 3 -r 11 -b 0 -c 64 -p} & 103.13 & 4.2h  & 522.9 & 87 \\
    & & Huge  & \texttt{-s 850 -i 3 -r 11 -b 0 -c 64 -p} & 120.1 & 4.3h & 627.3 & 87 \\
    \hline
    \multirow{3}{*}{AMG} &
    \multirow{3}{*}{\parbox{3.6cm}{Parallel algebraic multigrid solver
    for linear systems on unstructured grids. \textbf{FoM: $(nnz * (iters + steps))/seconds$}}}
    & Medium & \texttt{-problem 2 -n 340 340 340} & $5.66\mathrm{E}{8}$ & 7.7m & 72.2 & 209 \\
    
    & & Large & \texttt{-problem 2 -n 520 520 520} & $4.36\mathrm{E}{8}$ & 35.7m & 260.4 & 209 \\
    & & Huge & \texttt{-problem 2 -n 600 600 600} & $3.06\mathrm{E}{8}$ & 1.3h & 392.4 & 209 \\
    \hline
    \multirow{3}{*}{SNAP} &
    \multirow{3}{*}{\parbox{3.6cm}{Mimics the computational
    needs of PARTISN, a Boltzmann transport equation
    solver. \textbf{FoM: inverse of grind time (ns)}}}
    & Medium & nx=272, ny=102, nz=68 & $6.0\mathrm{E}{-2}$ & 12.9m & 61.4 & 87 \\
    & & Large  & nx=272, ny=272, nz=120 & $2.6\mathrm{E}{-2}$ & 2.3h & 288.8 & 90 \\
    & & Huge   & nx=272, ny=272, nz=192 & $2.4\mathrm{E}{-2}$ & 3.9h & 462.1 & 90 \\
    \hline
    \multirow{3}{*}{QMCPACK} &
    \multirow{3}{*}{\parbox{3.6cm}{Quantum Monte Carlo
    simulation of the electronic structure of atoms, molecules. \textbf{FoM: $(blocks*steps*N_w)/seconds$}}}
    & Medium & NiO S64 with VMC method, 40 walkers & $6.0\mathrm{E}{-2}$ & 10.2m & 16.5 & 1408 \\
    & & Large & NiO S128 with VMC method, 40 walkers & $1.3\mathrm{E}{-3}$ & 10.4h & 357.0 & 1402 \\
    & & Huge & NiO S256 with VMC method, 48 walkers & $3.3\mathrm{E}{-4}$ & 40.0h & 375.9 & 1408 \\
    \hline
  \end{tabular}
  \newline
  \vspace{1ex}
  \newline
  \begin{tabular}{|c|p{9.37cm}|r|r|r|r|}
    \hline
    \multicolumn{5}{|c|}{\textbf{SPEC\textsuperscript{\textregistered} CPU 2017}} \\
    \hline
    \multicolumn{1}{|c|}{\textbf{Application}}       &
    \multicolumn{1}{c|}{\textbf{Description}}        &
    \multicolumn{1}{c|}{\textbf{Time}}               &
    \multicolumn{1}{c|}{\textbf{GB}}                 &
    \multicolumn{1}{c|}{\textbf{Sites}}              \\
    \hline
    603.bwaves\_s
    & Numerically simulates blast waves in three dimensional transonic transient laminar viscous flow. 
    & 1.9m & 11.4 & 34 \\
    \hline
    607.cactuBSSN\_s
    & Based on Cactus Computational Framework, uses EinsteinToolkit to solve Einstein's equations in a vacuum. 
    & 2.7m & 6.6 & 809 \\
    \hline
    621.wrf\_s
    & Weather Research and Forecasting (WRF) Model, simulates one day of the Jan. 2000 North American Blizzard.
    & 3.1m & 0.2 & 4869 \\
    \hline
    627.cam4\_s
    & Community Atmosphere Model (CAM), atmospheric component for Community Earth System Model (CESM).
    & 7.6m & 1.2 & 1691 \\
    \hline
    628.pop2\_s
    & Parallel Ocean Program (POP), simultaneously simulates earth's atmosphere, ocean, land surface and sea-ice.
    & 3.6m & 1.5 & 1107 \\
    \hline
    638.imagick\_s
    & Performs various operations to transform an input image and compares the result to a reference image.
    & 5.4m & 6.9 & 4 \\
    \hline
    644.nab\_s
    & Nucleic Acid Builder (NAB), performs FP calculations that occur commonly in life science computation.
    & 3.2m & 0.6 & 88 \\
    \hline
    649.fotonik3d\_s
    & Computes transmission coefficient of a photonic waveguide using the FDTD method for Maxwell's equations.
    & 3.2m & 9.5 & 127 \\
    \hline
    654.roms\_s
    & Regional Ocean Modeling System, forecasts water temperature, ocean currents, salinity, and sea surface height.
    & 4.9m & 10.2 & 395 \\
    \hline
  \end{tabular}
  \label{tab:benchmarks}
\end{table*}
\vspace{-1ex}
\subsection{Workloads}
Our evaluation employs applications from two popular sets of benchmark programs: CORAL~\cite{coral}, which includes several widely used HPC applications and proxy applications, and SPEC\textsuperscript{\textregistered} CPU 2017~\cite{SPECCPU17}, which is comprised of a variety of industry standard applications for stressing processor and memory performance.
From CORAL, we selected three proxy applications (LULESH, AMG, and SNAP) and one full scale scientific computing application (QMCPACK) based on their potential to stress cache and memory performance on our platform.
%as well as our own prior experience building and running these benchmarks.
To study the impact of online tiering guidance with more varied inputs and capacity requirements, we also constructed and evaluated three separate input sizes for each CORAL application.
The top part of Table~\ref{tab:benchmarks} provides descriptions and relevant usage statistics for each CORAL application-input pair included in this study.

The benchmarks in SPEC\textsuperscript{\textregistered} CPU 2017 are designed to test a variety of system behavior, and include several single-threaded and CPU-bound applications as well as memory intensive programs.
For this study, we focused our evaluation on those floating point (FP) benchmarks that provide the option to distribute their processing over a configurable number of application threads through the use of OpenMP directives.\footnote{Specifically, our study includes all FP benchmarks in SPEC\textsuperscript{\textregistered} CPU 2017 with OpenMP directives with the exception of \emph{619.lbm\_s}, which is omitted because it only allocates a single, large heap object throughout its entire run, and is therefore not likely to exhibit benefits with guided object placement.}
When configured to use larger numbers of software threads, these FP workloads tend to have relatively high memory bandwidth requirements, and thus, magnify the importance of data placement on our platform.
The bottom part of Table~\ref{tab:benchmarks} provides descriptions and other relevant usage statistics for our selected SPEC\textsuperscript{\textregistered} CPU benchmarks.
All of our experiments with these benchmarks use the standard \emph{ref} program input.

\vspace{-1ex}
\subsection{Common Experimental Configuration}
All applications were compiled using the LLVM compiler toolchain (v. 7.1.0) with default optimization settings and \texttt{-march=x86\_64}.
%All used their default optimization settings for each application with
%all configurations.
C/C++ codes use the standard \texttt{clang} frontend, and Fortran codes are converted to LLVM IR using Flang~\cite{flang}.
All guided and non-guided configurations use SICM with the unmodified \texttt{jemalloc} allocator (v. 5.2.0) with \texttt{oversize\_threshold} set to 0, \texttt{background\_thread} set to \texttt{true}, and \texttt{max\_background\_threads} set to 1.\footnote{Setting \texttt{oversize\_threshold} to 0 disables a feature of \texttt{jemalloc} that allocates objects larger than a specific size to a dedicated arena (to reduce fragmentation). The other two parameters control the number of background threads, which enable \texttt{jemalloc} to purge unused pages asynchronously.}
To prepare executables for guided execution, we configure the compilation pass to clone up to three layers of call path context to each allocation site.
Our previous work has shown that this amount of context is sufficient to obtain the benefits of this approach for most applications~\cite{olson2018nas,effler2019memsys}.

For the default and offline configurations, each benchmark is configured to use 16 software threads to match the number of cores on our experimental platform.
The offline configuration always uses the same program input for the profile and evaluation runs.
The online configuration, as well as the profile run of the offline configuration, only create 15 software threads for each application because they require an additional runtime thread to profile and periodically enforce data-tier recommendations.
We tested the alternative strategy of over-provisioning compute resources by running 16 application threads alongside this extra thread, and leaving it to the system scheduler to resolve conflicts for computing cores.
However, we found that this approach consistently produced much worse performance than the 15-thread configuration with our benchmarks.

To reduce sources of variability between runs, all of our experiments execute each application in isolation on an otherwise idle machine.
Prior to each experimental run, an automated script clears out the Linux page cache and disables transparent huge pages for the application process.

To estimate the usage rate of each site, the offline and online profilers use the Linux perf~\cite{perf} facility to sample memory reads from the target application that miss the last level processor caches.
Specifically, we sample \texttt{MEM\_LOAD\_L3\_MISS\_RETIRED} event on our platform with a PEBS reset value of 512.
We also compute the resident set size for each site by counting the number of active (4 KB) pages associated with the site's corresponding VMA, as described in Section~\ref{sec:rss}.

For the online approach, we experimented with a number of interval lengths for analyzing profile information and migrating program data (i.e., the \emph{IntervalTime} parameter in Algorithm~\ref{alg:online}), including: 0.1s, 1s, 10s, and 100s.
We found that relatively short intervals of 1s or less were more sensitive to shifts in memory usage, but also incurred higher overheads due to more frequent interruptions and data migrations.
Of the interval lengths we tested, 10s provided the best balance of relatively low migration overheads with relatively quick convergence to a good data-tiering configuration, and provided the best overall performance for the applications we tested.
Hence, all of our online results in the next section use an interval length of 10s.

Additionally, we configured the hybrid arena allocator to promote an allocation context to its own thread-shared arena after it allocates more than 4 MB of data (in total) to the thread-private arenas.
With this configuration, all of our benchmarks allocate the vast majority of their data objects to the shared arenas.
Specifically, the peak capacity of the private arenas is no more than a few MBs in all but two benchmarks (621.wrf\_s and 627.cam4\_s).
In the worst case of 627.cam4\_s, the peak RSS of the private arenas is 0.3 GBs.

Lastly, in previous works with offline guided data tiering, we have found that the \emph{thermos} approach is the most effective approach for converting profile information to memory tier recommendations~\cite{olson2018nas,olson2019memsys}.
Hence, in this work, all of the offline and online guided data tiering configurations use \emph{thermos} to partition the allocation sites into sets for the faster and slower memory tiers.

\vspace{-1ex}
\subsection{Reporting Details}
Aside from the results showing profile overhead, which use execution time, all performance measurements for each configuration and application are presented as throughput.
For the CORAL benchmarks, we report the application-specific figure of merit (FoM), which, for our selected benchmarks, is always a measure of throughput.
For SPEC\textsuperscript{\textregistered} CPU 2017, we report time per operation (i.e., the inversion of wall clock execution time) for each benchmark run.

Except for the CORAL benchmarks with large and huge input sizes, all results are reported as the mean average of five experimental runs relative to the default configuration.
The larger CORAL inputs often require multiple hours for even a single run, and so we only conducted one run of each to limit computing time.
For experiments with multiple runs, we estimate variability and significance of our results by computing the 95\% confidence intervals for the difference between the mean results of the experimental and default configurations, as described in~\cite{georges2007oopsla}.
These intervals are plotted as error bars around the sample means in the relevant figures.
However, it is important to note that variability is too low for these bars to be visible in some figures.

%%%%%%%%%%%%%%%%%%%%%%%%%%%%%%%%%%%%%%%%%%%%%%%%%%%%%%%%%%%%
% RESULTS
%%%%%%%%%%%%%%%%%%%%%%%%%%%%%%%%%%%%%%%%%%%%%%%%%%%%%%%%%%%%

\iffalse
\begin{figure}[t]
    \begin{minipage}[t]{0.49\textwidth} %
        \includegraphics[width=\textwidth]{figures/profile_coral_medium.pdf} %
    \end{minipage} %
    \begin{minipage}[t]{0.49\textwidth} %
        \includegraphics[width=\textwidth]{figures/profile_coral_large.pdf} %
    \end{minipage} %
    \label{fig:profile_overhead_coral}
    \caption{...}
\end{figure}
\fi

\vspace{-1ex}
\section{Evaluation}
\subsection{Online Profile Overhead}
Let us first consider the performance overhead of collecting memory usage information during program execution.
For this evaluation, we compare the previous offline profiling approach and our online profiling enhancements described in Section~\ref{sec:online_profiling} (but without any data migration mechanisms), to a default configuration.
In the default configuration, each benchmark program is configured to use 16 application threads (one for each hardware thread on our evaluation platform) and the unmodified jemalloc allocator with default execution options.
In contrast, each profile configuration uses only 15 program threads and an extra thread to conduct all profiling operations.
For brevity, this section omits results for the CORAL benchmarks with the large and huge input sizes.
% MBO: In the previous sentence, if you want, we can use the following metrics in
% order to alleviate concerns that we're not including `large` and `huge` profiling
% overhead due to poor results:
% 1. For CORAL `large` runs, the new profiling incurs a 1.7% overhead over a default
%    configuration.
% 2. For CORAL `huge`, which uses the DRAM as a cache for the Optane DC memory due
%    to its large size, the geometric mean of the performance overhead is 10.6%.

\begin{figure}[t]
  \centering
  \includegraphics[width=0.85\linewidth]{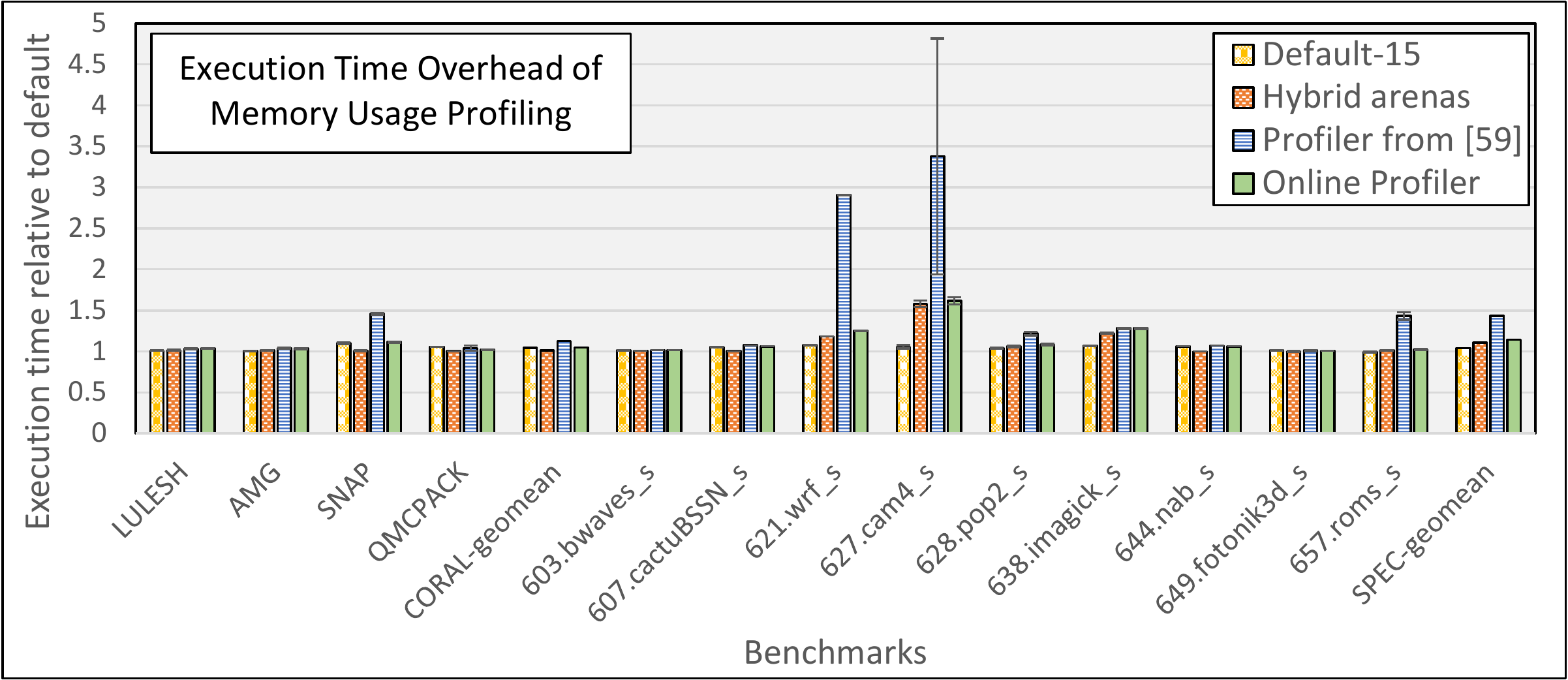}
  \captionof{figure}{Execution time overhead of memory usage profiling (lower is better).}
  \label{fig:profile_overhead}
\end{figure}
Figure~\ref{fig:profile_overhead} shows the execution time overhead of the previous offline and current online profiling mechanisms relative to the default configuration.
In addition to the two profiling configurations, we also tested two other configurations to isolate the impact of using one less application thread as well as the effect of the hybrid arena allocation strategy described in Section~\ref{sec:hybrid_arenas}.
The \emph{default-15} configuration is identical to the baseline, but uses only 15 application threads.
\emph{Hybrid arenas} uses our custom allocator to apply the hybrid arena allocation strategy, and also uses only 15 application threads, but does not collect any profile information.

The results show that using one less application thread to reserve one core for profiling only has a marginal impact for these workloads on our 16-core processor.
Overall, both the CORAL and SPEC\textsuperscript{\textregistered} sets run about 4\% slower with one less thread, on average, with a worst case slow down of 10\% for SNAP.
The hybrid arena allocator has a mixed effect. In some cases, such as SNAP and QMCPACK, this approach actually improves performance over the default allocator.
However, for some SPEC\textsuperscript{\textregistered} benchmarks, such as 627.cam4\_s and 638.imagick\_s, it can cause significant slow downs.
On average, and in comparison to the \emph{default-15} configuration, the hybrid allocator improves performance by 3\% for the CORAL benchmarks, and degrades performance by 6\% for SPEC\textsuperscript{\textregistered}.

% MBO: When you say "the previous offline profiling approach," will that make future
% evaluation sections confusing, because we're going to include an offline
% approach (but not the same profiling)?

\begin{wraptable}[17]{r}{8.2cm}
  \vspace{-2ex}
  \centering
  \footnotesize
  \begin{tabular}{|l|c|c|c|c|}
    \hline
    \multicolumn{1}{|c|}{\multirow{2}{*}{Application}} & \multicolumn{2}{|c|}{Mean Profile Time (s)} & \multicolumn{2}{|c|}{Max Profile Time (s)} \\
    \cline{2-5}
    & Offline~\cite{olson2019memsys} & Online & Offline~\cite{olson2019memsys} & Online \\
    \hline
    LULESH & 1.57 & 0.754 & 4.303 & 2.834 \\
    AMG & 1.345 & 0.356 & 3.043 & 1.227 \\
    SNAP & 1.313 & 0.413 & 3.141 & 1.178 \\
    QMCPACK & 20.006 & 0.561 & 57.101 & 1.294 \\
    \textbf{CORAL Mean} & \textbf{6.058} & \textbf{0.521} & \textbf{16.897} & \textbf{1.633} \\
    \hline
    603.bwaves\_s & 0.372 & 0.134 & 0.426 & 0.166 \\
    607.cactuBSSN\_s & 1.886 & 0.031 & 25.548 & 0.525 \\
    621.wrf\_s & 6.83 & 0.102 & 8.45 & 0.12 \\
    627.cam4\_s & 0.535 & 0.018 & 0.984 & 0.024 \\
    628.pop2\_s & 3.032 & 0.132 & 5.021 & 0.162 \\
    638.imagick\_s & 0.038 & 0.001 & 0.073 & 0.009 \\
    644.nab\_s & 0.113 & 0.097 & 0.349 & 0.325 \\
    649.fotonik3d\_s & 0.833 & 0.431 & 1.918 & 0.787 \\
    654.roms\_s & 6.709 & 0.904 & 15.4 & 2.763 \\
    \textbf{SPEC\textsuperscript{\textregistered} Mean} & \textbf{2.260} & \textbf{0.205} & \textbf{6.463} & \textbf{0.542} \\
    \hline
  \end{tabular}
  \caption{Mean and maximum time (in seconds) to collect a single profile using the offline profiler from \cite{olson2019memsys} vs. our online profiler.}
  \label{tab:interval_time}
\end{wraptable}
We also find that the new online profiler, which includes the hybrid arena allocator and more efficient RSS accounting, significantly outperforms the previous offline profiling approach.
On average, execution with the online profiler is 8\% faster with the CORAL benchmark set and 26\% faster with SPEC\textsuperscript{\textregistered} compared to the offline profiling approach.
Relative to the default configuration with no online profiling and an extra application thread, the online profiler adds 5\% and 14\% execution time overhead for CORAL and SPEC\textsuperscript{\textregistered}, respectively.
If the system includes at least one free computing core to run the profile thread, the overhead is even lower.
Specifically, in comparison to the \emph{default-15} configuration, the online profiler adds $<$1\% and $<$10\% execution time cost, on average, for the CORAL and SPEC\textsuperscript{\textregistered} benchmark sets, respectively.

% MBO: Not sure why you wanted to exclude this paragraph; I like how it's phrased
% both in your Discord message and here (although it sounds more polished here).

In most cases, the execution time cost is due to the use of the alternative arena allocation strategy during profiling.
Indeed, comparing the \emph{online profiler} and \emph{hybrid arenas} configurations directly shows that enabling the memory access and RSS tracking adds only about 3.5\% overhead, on average, across all of our selected benchmarks.
Hence, while the overhead for profiling is already relatively low compared to previous works, further optimization efforts that enable the runtime to compute and organize the necessary information for each allocation context without affecting data locality could reduce it even further.

\vspace{-1ex}
\paragraph{Time per Profile Interval}
In addition to the overall performance improvements shown in Figure~\ref{fig:profile_overhead}, the enhancements developed for this work also reduce the length of time necessary to collect a full memory usage profile.
These reductions enable the runtime to make faster and more effective tiering decisions based on more recent memory behavior.
Table~\ref{tab:interval_time} shows the mean average and maximum number of seconds necessary to collect and analyze profiles of every allocation context during program execution for all of the selected benchmarks.
We find that the enhancements substantially reduce the time per profile interval.
On average, profile interval time is reduced by more than $11x$ across both benchmark sets.
The vast majority of this improvement is driven by the system-level instrumentation that tracks the number of pages mapped for each arena.
Since it is no longer necessary to count the number of pages mapped for each arena during each profile interval, the profiler can estimate the capacity of each allocation context much more quickly.

\vspace{-1ex}
\subsection{Performance of Guided Data Management with Varying Capacity Constraints}
\begin{figure*}
%\captionsetup{skip=4pt}
\centering
\subfloat[LULESH\label{medium_lulesh}]{
  \includegraphics[width=0.32\columnwidth]{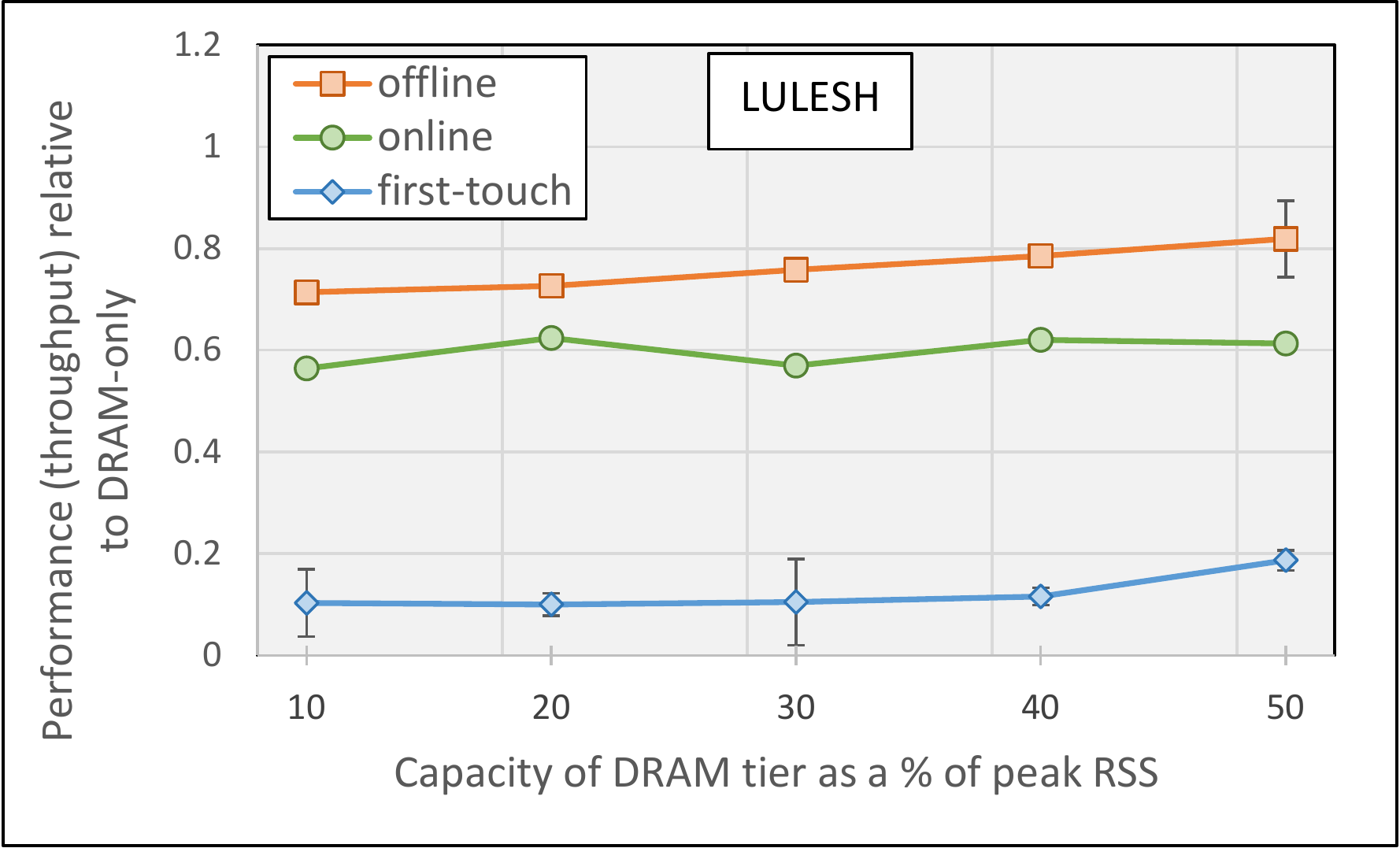}
}
\hfill%
\subfloat[AMG\label{medium_amg}]{
  \includegraphics[width=0.32\columnwidth]{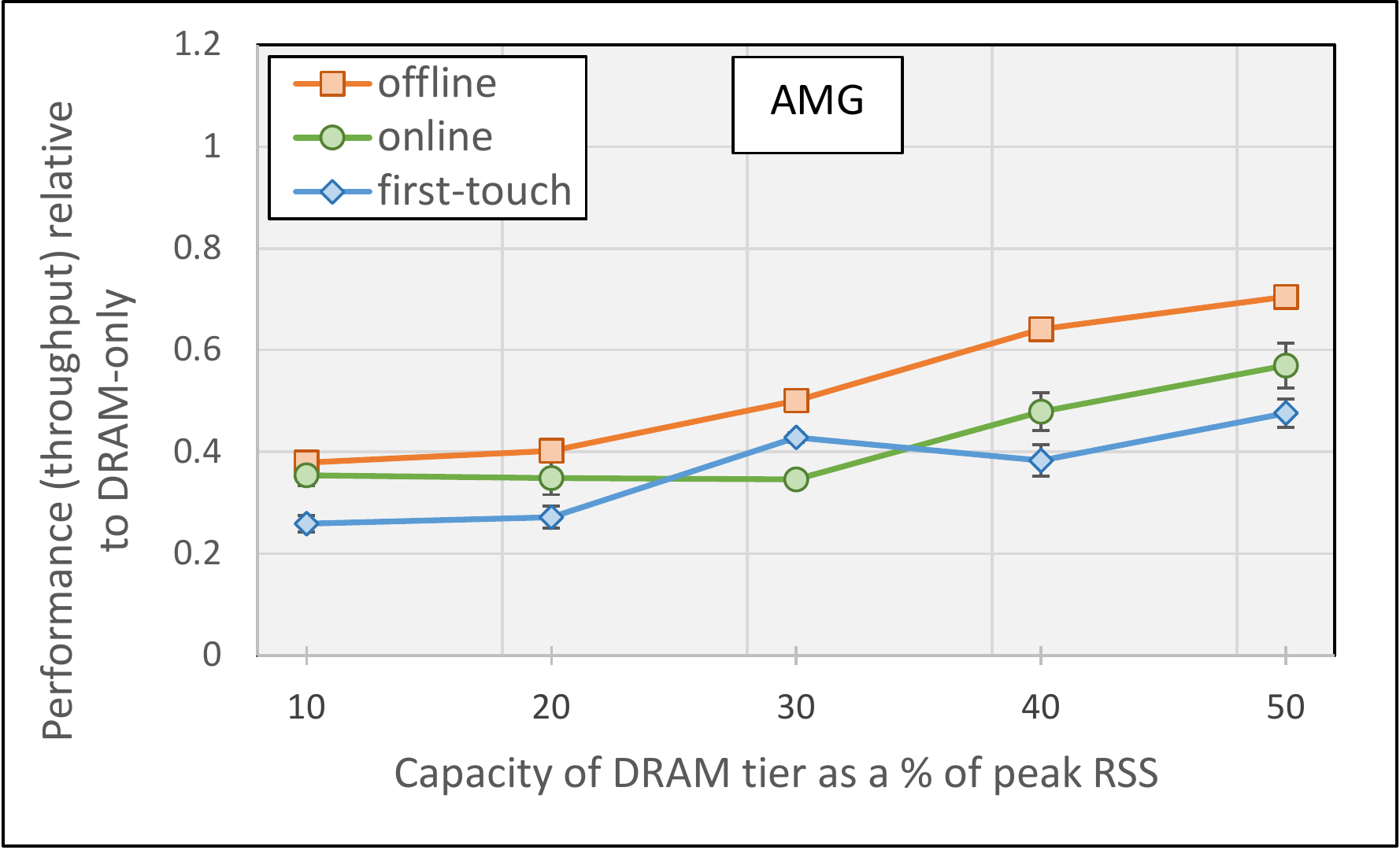}
}
\hfill%
\subfloat[SNAP\label{medium_snap}]{
  \includegraphics[width=0.32\columnwidth]{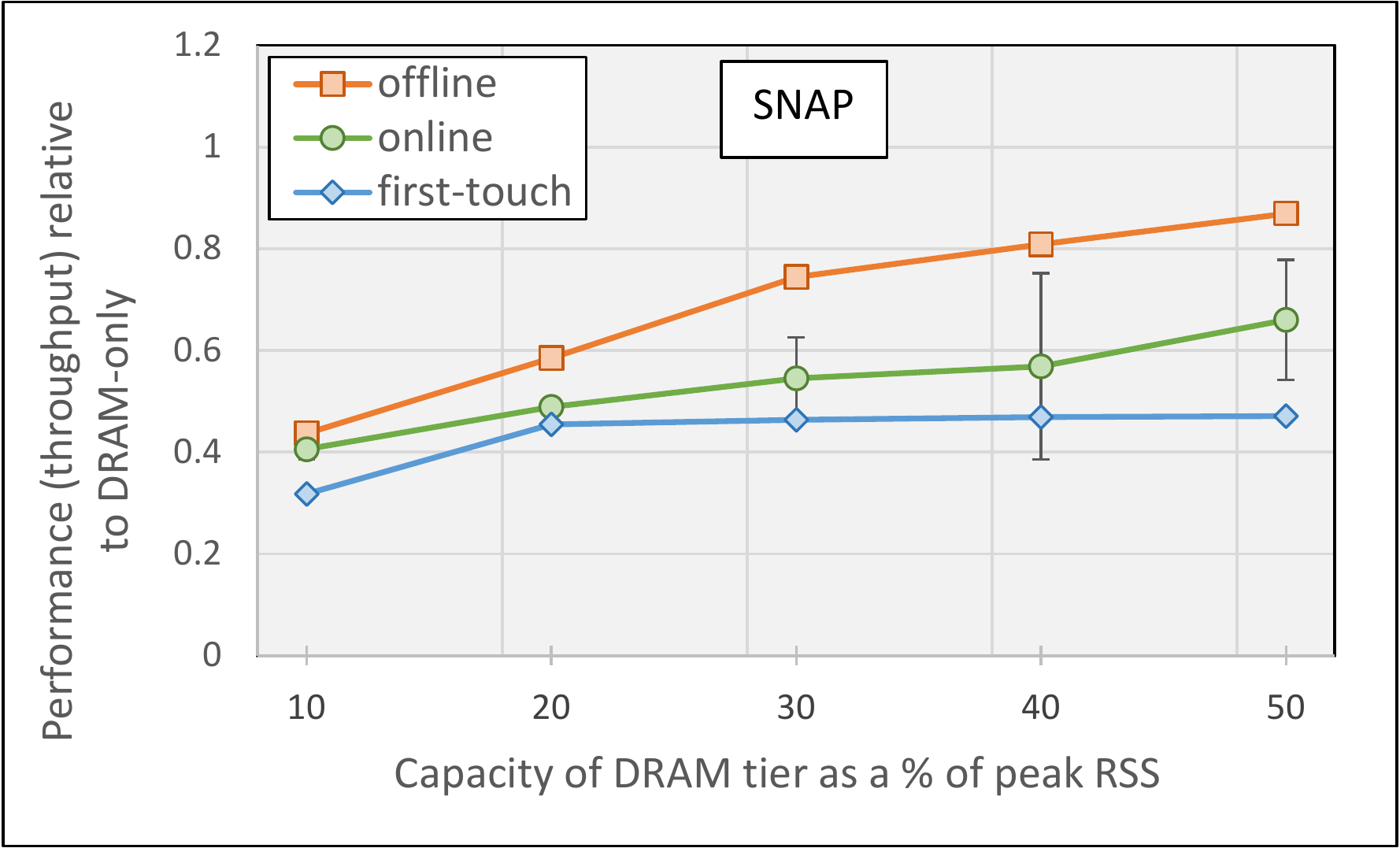}
}\\
\subfloat[QMCPACK\label{medium_qmcpack}]{
  \includegraphics[width=0.32\columnwidth]{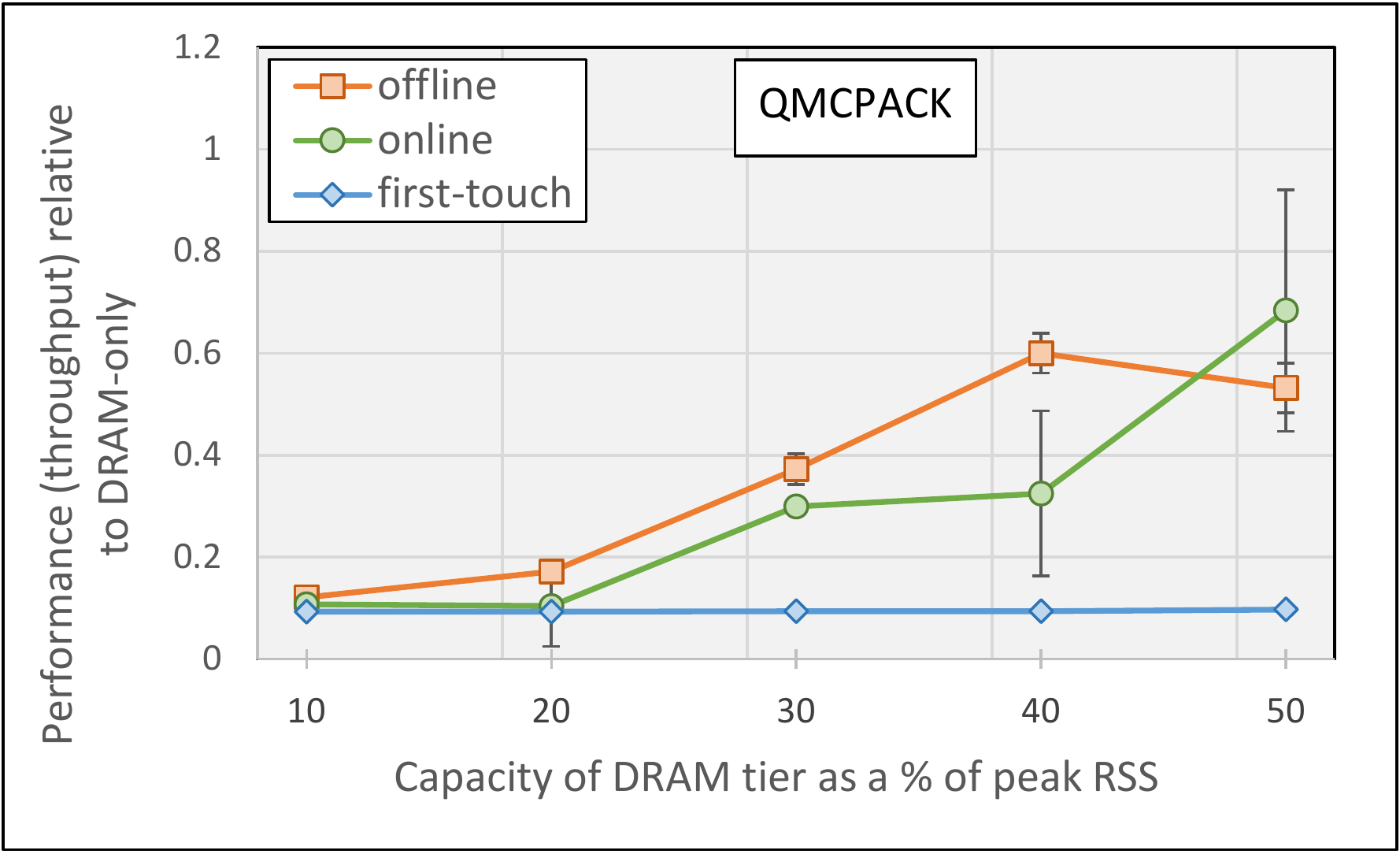}
}
\hfill%
\subfloat[603.bwaves\_s\label{ref_bwaves}]{
  \includegraphics[width=0.32\columnwidth]{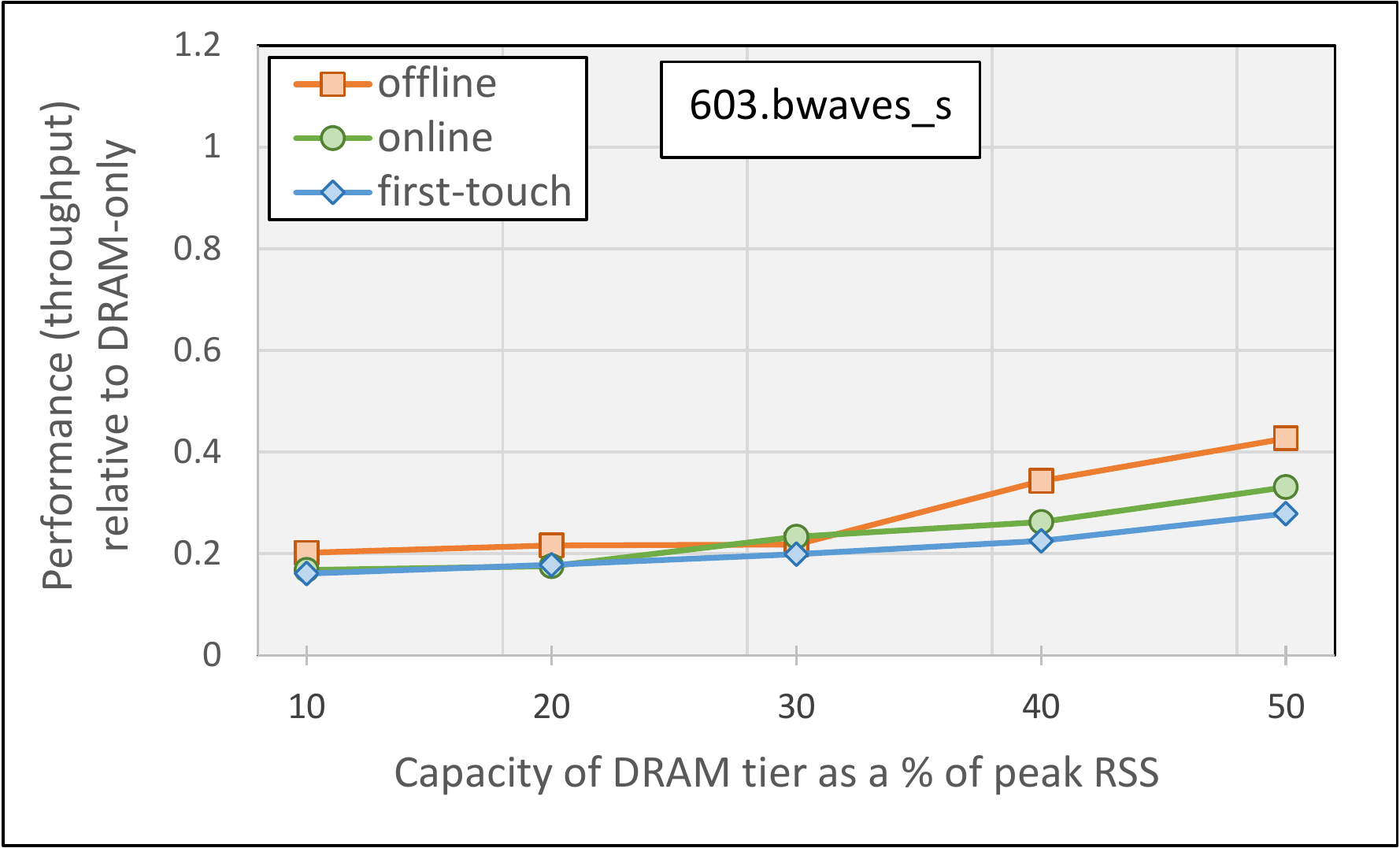}
}
\hfill%
\subfloat[607.cactuBSSN\_s\label{ref_cactuBSSN}]{
  \includegraphics[width=0.32\columnwidth]{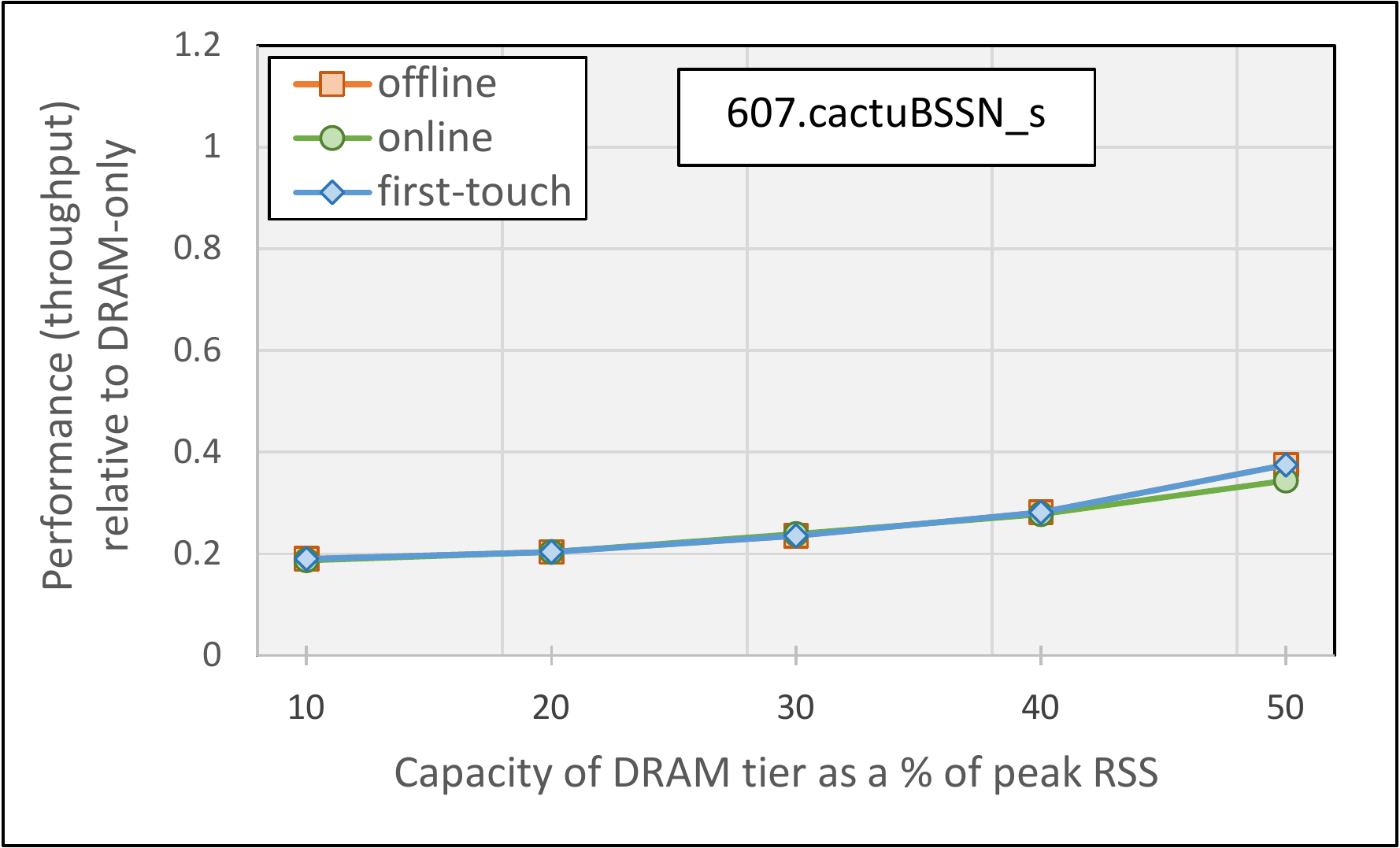}
}\\
\subfloat[621.wrf\_s\label{ref_wrf}]{
  \includegraphics[width=0.32\columnwidth]{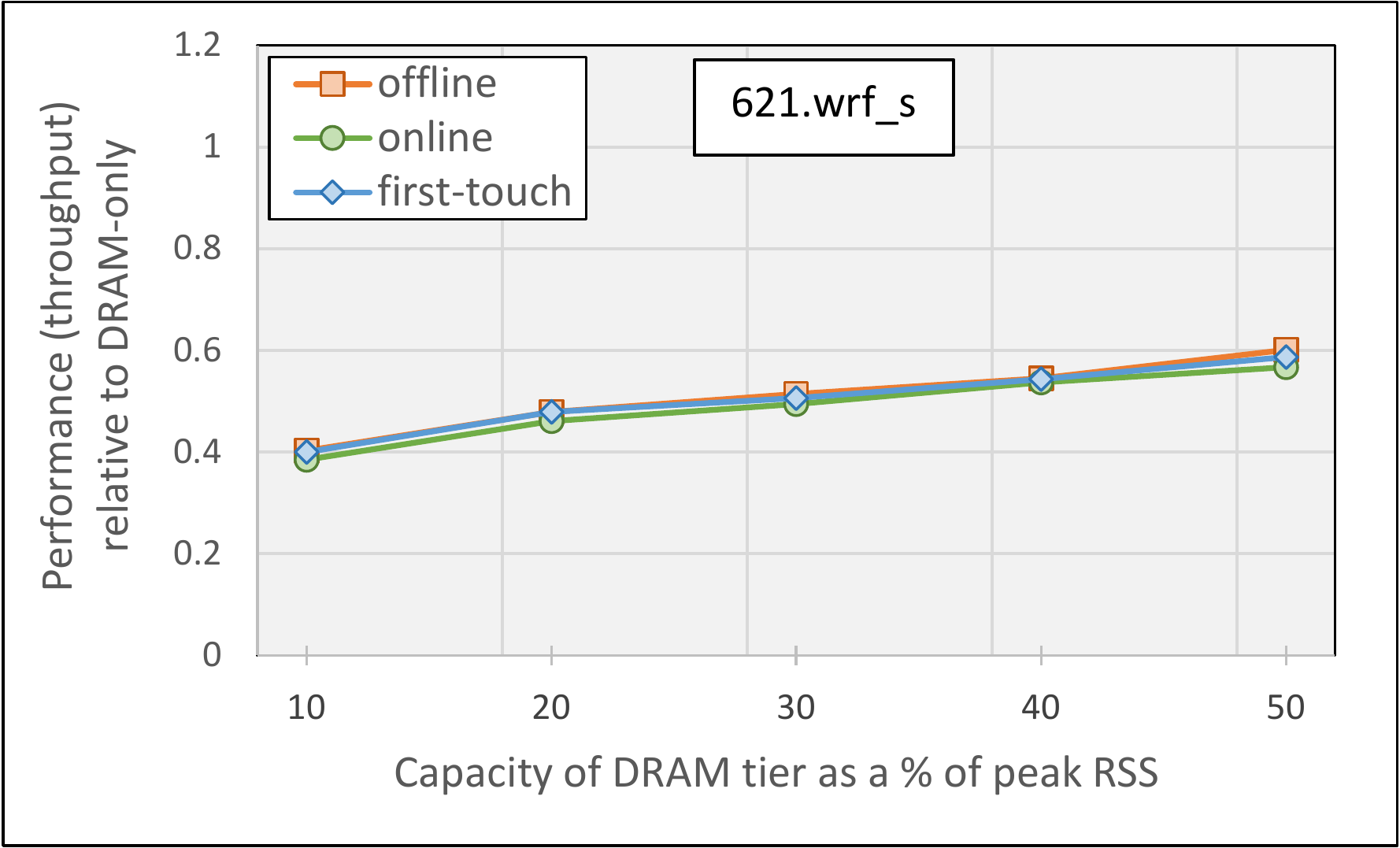}
}
\hfill%
\subfloat[627.cam4\_s\label{ref_cam4}]{
  \includegraphics[width=0.32\columnwidth]{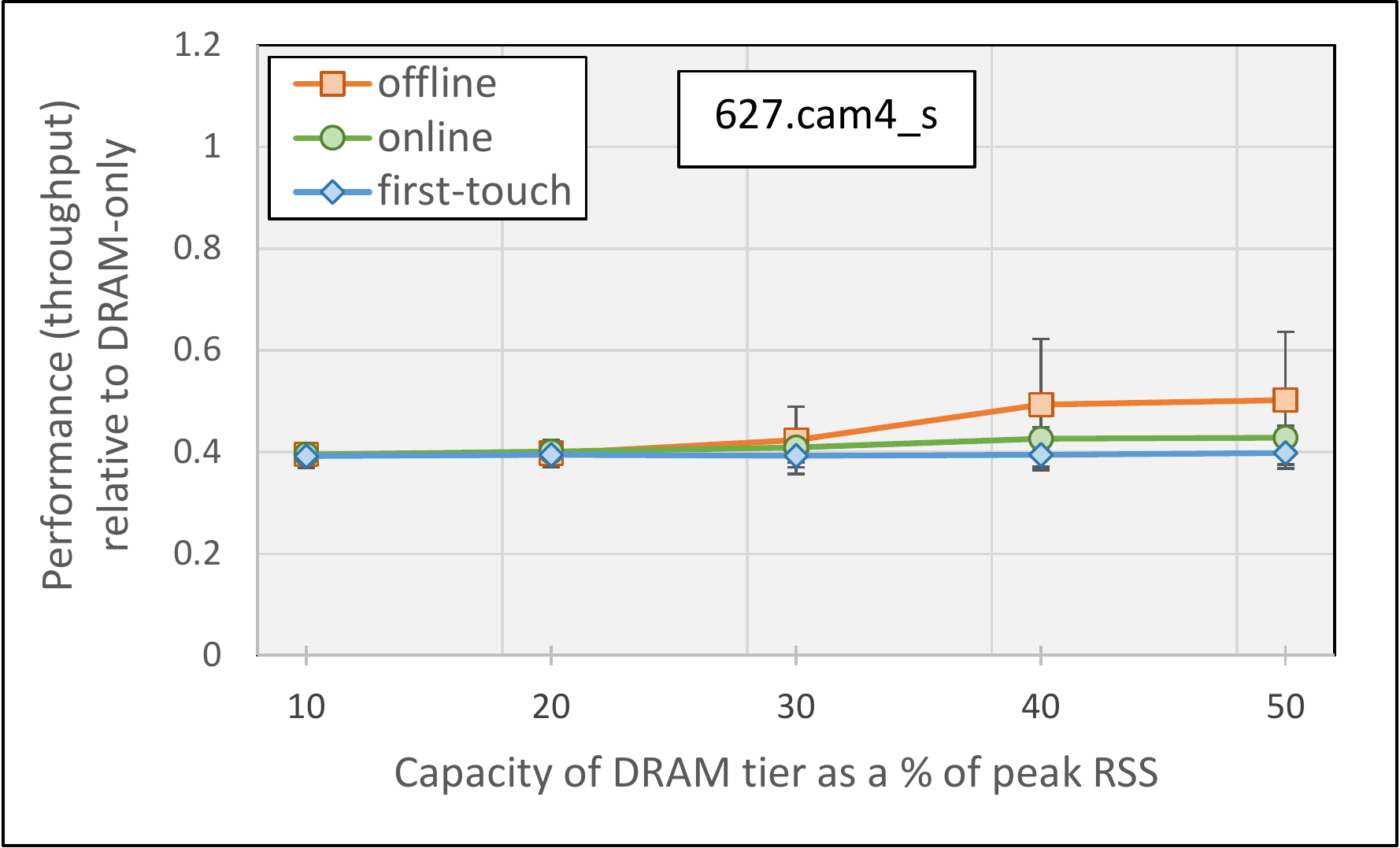}
}
\hfill%
\subfloat[628.pop2\_s\label{ref_pop2}]{
  \includegraphics[width=0.32\columnwidth]{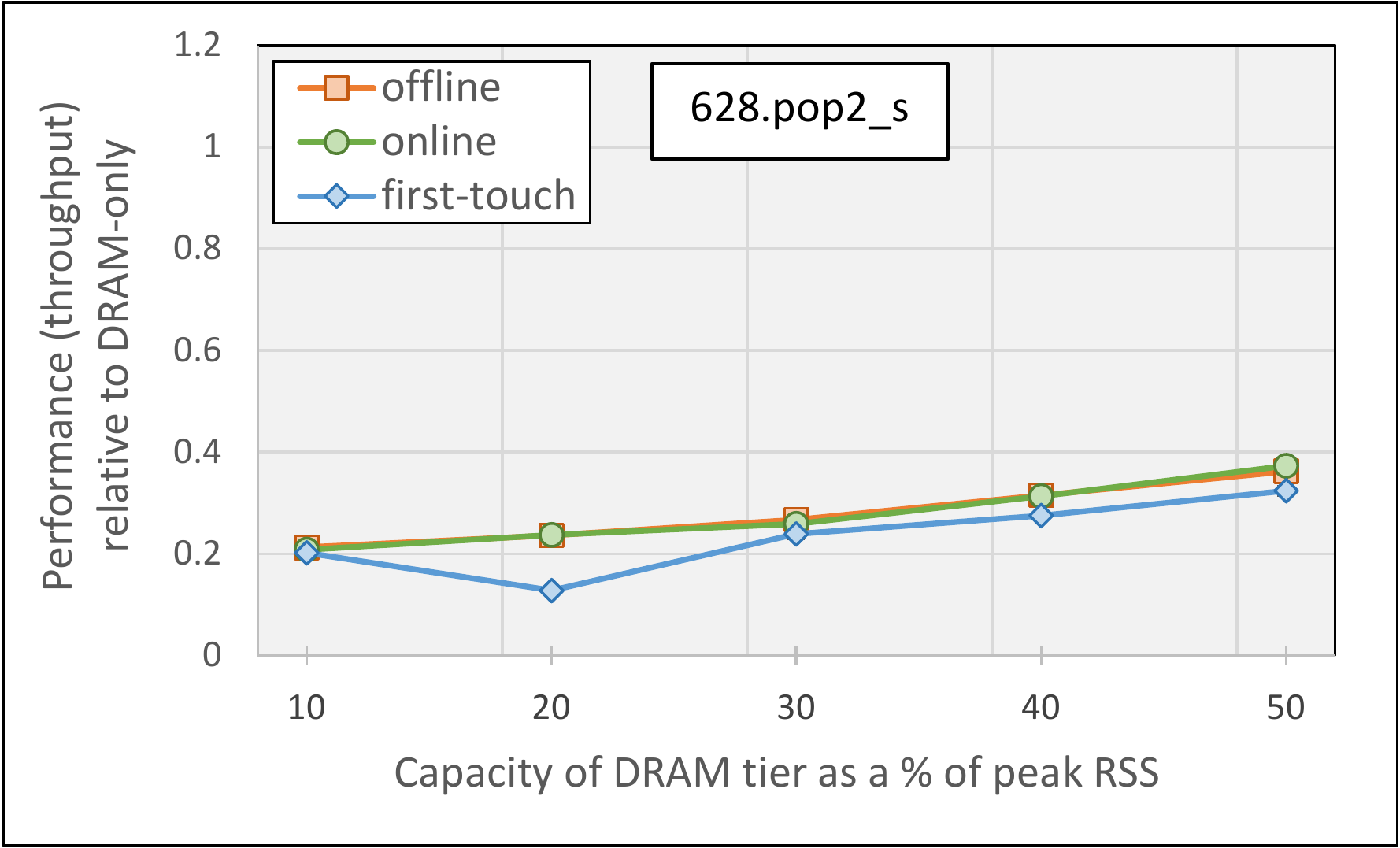}
}\\
\subfloat[638.imagick\_s\label{ref_imagick}]{
  \includegraphics[width=0.235\columnwidth]{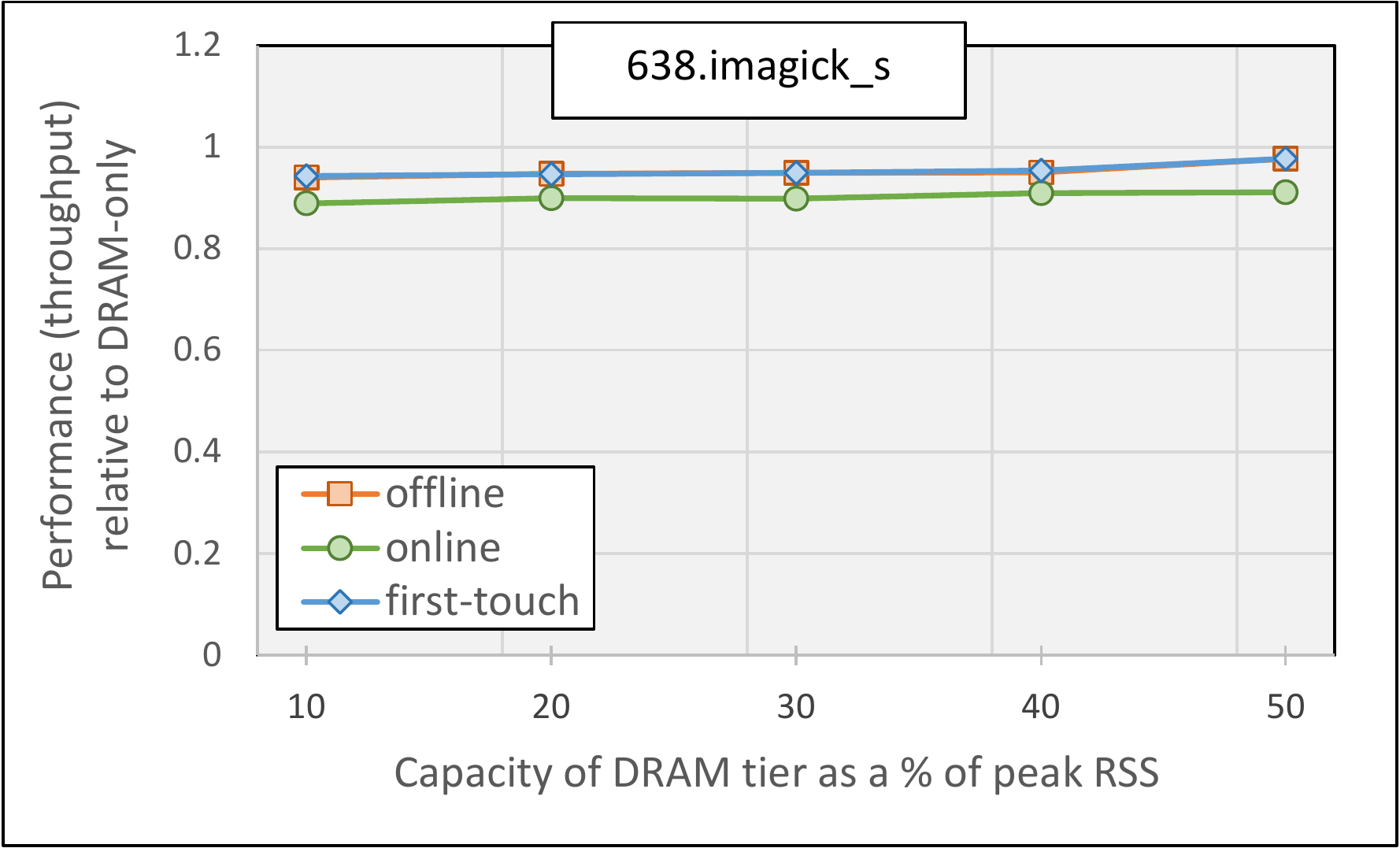}
}
\hfill%
\subfloat[644.nab\_s\label{ref_nab}]{
  \includegraphics[width=0.235\columnwidth]{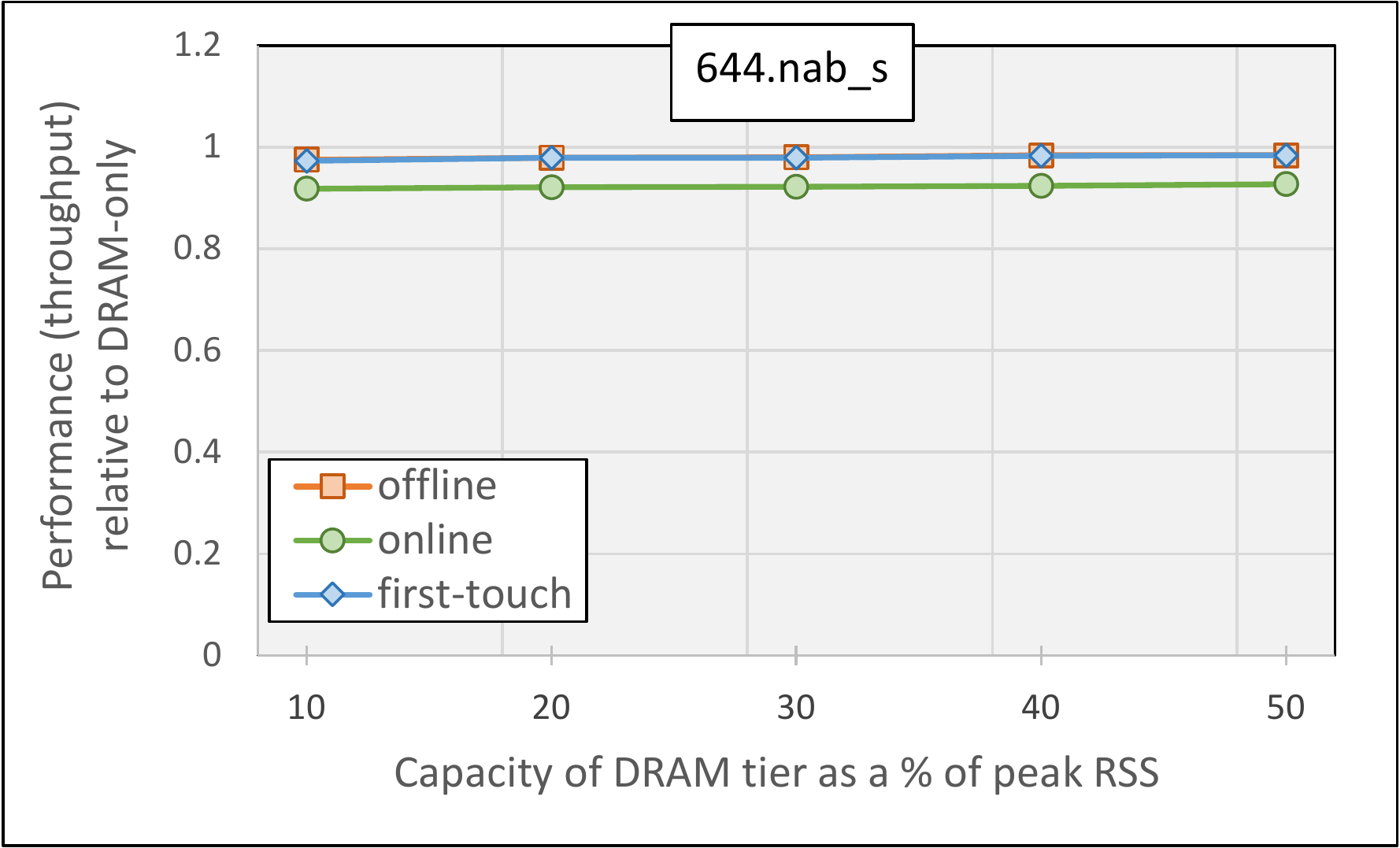}
}
\hfill%
\subfloat[649.fotonik3d\_s\label{ref_fotonik3d}]{
  \includegraphics[width=0.235\columnwidth]{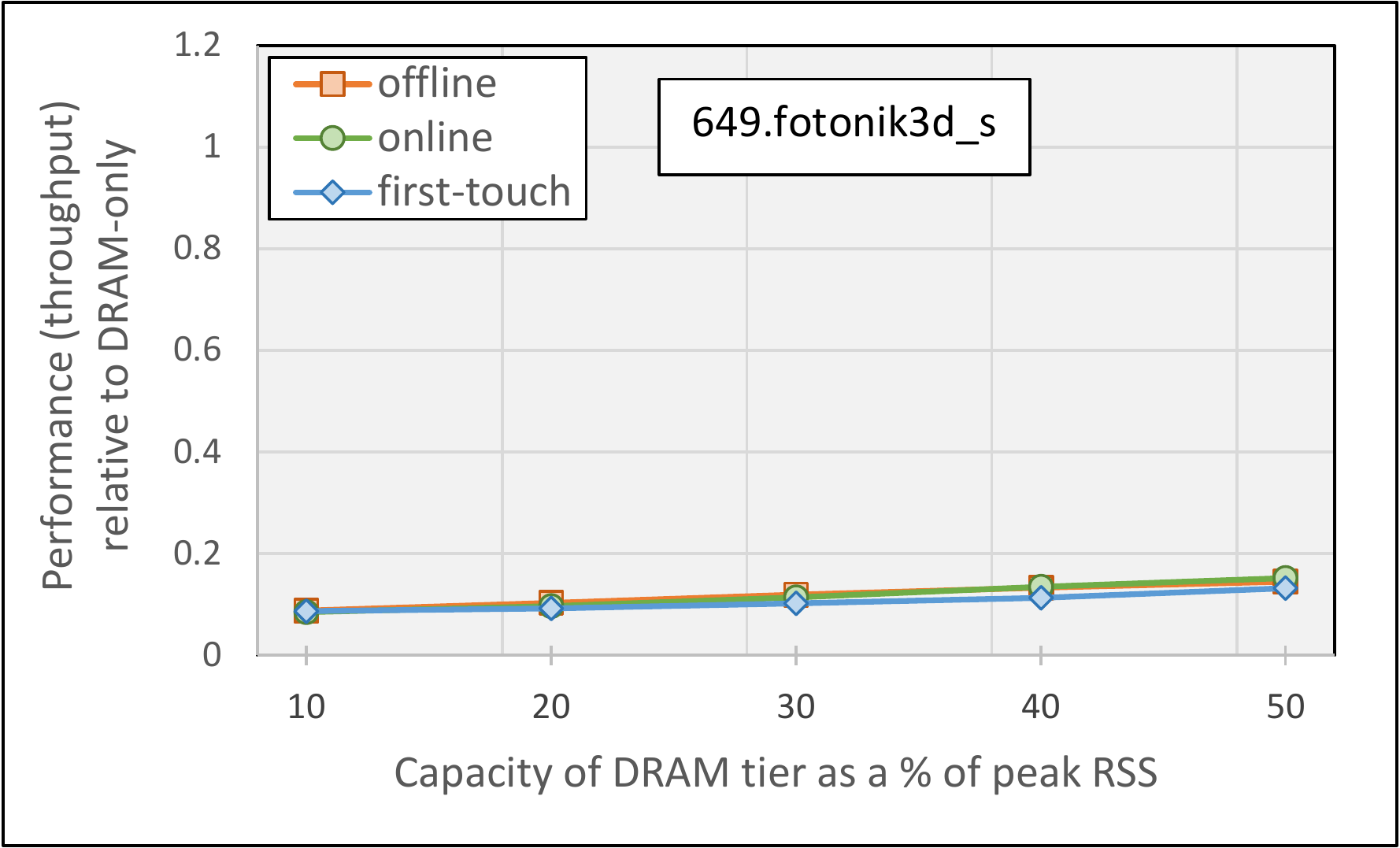}
}
\hfill%
\subfloat[654.roms\_s\label{ref_roms}]{
  \includegraphics[width=0.235\columnwidth]{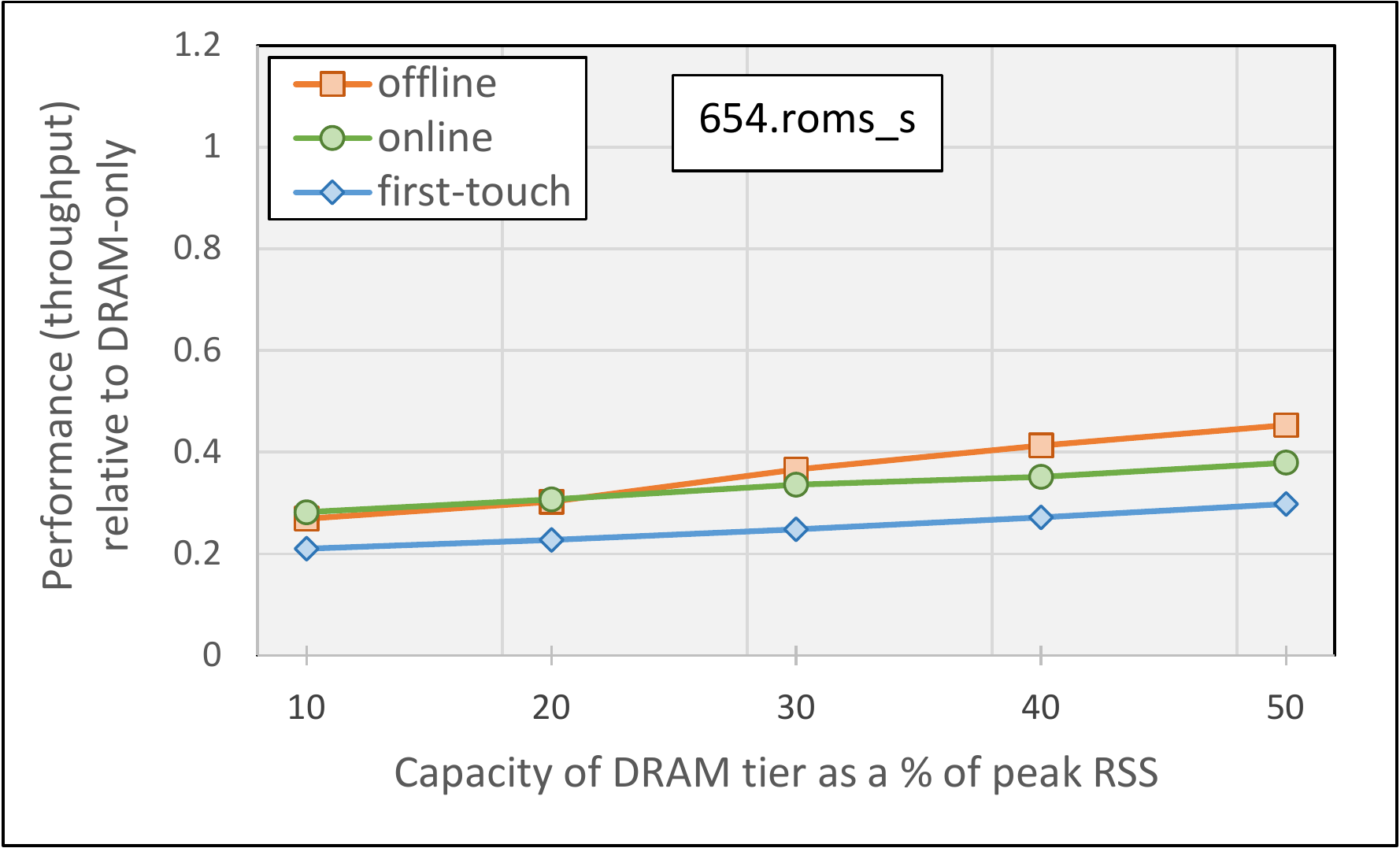}
}
\caption{Performance (throughput) of offline and online guidance based approaches, compared to the unguided first touch configuration, with varying amounts of capacity available in the faster DDR4 memory tier. All results are shown relative to a configuration with all program data allocated to the DDR4 tier (higher is better). Note that the DDR4 tier capacities shown along the x-axis are calculated as a percentage of the peak resident set size during execution with the default configuration.}
\label{fig:medium_performance}
\end{figure*}

Our next set of experiments aim to evaluate the performance of offline and online guided data management with varying capacity constraints in the faster memory tier. 
For this evaluation, we again use the CORAL benchmarks with medium input sizes as well as the selected SPEC\textsuperscript{\textregistered} CPU 2017 benchmarks.
To evaluate each workload and configuration with different capacity constraints, we extended our Linux kernel with new facilities to control the amount of DRAM available for a given process.
Specifically, we added an option to the memory control group (cgroup) interface~\cite{memcg} to allow individual processes or process groups to limit the amount of physical memory that they are able to allocate and keep resident on a particular NUMA node at any given point in time.
Thus, if a process attempts to map a virtual page to a new physical page on a node whose specified limit has already been reached, the kernel will force the process to use a page from a different NUMA node to satisfy the fault, or fallback to page reclaim if no other memory is available.

To prepare these experiments, we first measured the peak resident set size of a run of the default configuration of each benchmark application.
Subsequent experiments then use the cgroup controls to limit the capacity available in the DRAM tier to be a percentage of the measured peak RSS of the running application.
Specifically, we tested configurations with DRAM capacity limited to 10\%, 20\%, 30\%, 40\%, and 50\% of the peak RSS of the application.
For comparison against a standard data tiering approach that does not use any profile guidance, we also ran each benchmark with each capacity limit with an unguided \emph{first touch} configuration.
The first touch configuration simply satisfies all memory demands from the application with allocations from the DRAM tier if capacity is available, and otherwise, from the Optane\textsuperscript{TM} tier.

Figure~\ref{fig:medium_performance} presents the performance of each benchmark with first touch as well as the offline and online guided data tiering approaches.
All results show throughput relative to the default configuration with no capacity limitations in the DRAM tier (i.e., all memory objects use the faster memory devices), and thus, higher is better.
We can make several key observations based on these results.
First, profile guided data tiering enables significant speedups compared to first touch for all four CORAL benchmarks.
In the best cases, the offline approach is up to $7.3x$ faster than first touch (LULESH, 20\% DRAM), while the online approach is up to $7.1x$ faster (QMCPACK, 50\% DRAM).
Average (geometric mean) speedups with the CORAL benchmarks range from $2.1x$ to $3.3x$ for the offline approach, and $1.8x$ to $2.5x$ for the online approach, across the different capacity limits.

The performance impact of profile guided data tiering with the SPEC\textsuperscript{\textregistered} benchmark set is more modest, but still significant.
Several benchmarks (specifically, 607.cactuBSSN\_s, 621.wrf\_s, 638.imagick\_s, and 644.nab\_s) exhibit little or no improvement with guided data management.
In some cases (specifically, 638.imagick\_s, and 644.nab\_s), the online approach actually slightly degrades performance because the overhead of profiling is not offset by any gains in efficiency.
However, guided data tiering does enable significant speedups for the other SPEC\textsuperscript{\textregistered} benchmarks.
For instance, the offline approach speeds up some configurations of 603.bwaves\_s and 654.roms\_s by more than 50\%, while the online approach speeds up these applications by up to 18\% and 35\%, respectively, compared to first touch.
The best case for both the offline and online approaches is 628.pop2\_s with 20\% DRAM capacity, which speeds up by more than 84\% with either guided approach.
Overall, and across the different capacity limits, average speedups for the full group of SPEC\textsuperscript{\textregistered} benchmarks ranges from $5.7\%$ to $14.6\%$ for the offline approach and $1.8\%$ to $8.6\%$ for the online approach.

\begin{figure*}
%\captionsetup{skip=4pt}
\centering
\subfloat[LULESH\label{fig:lulesh_bandwidth_intervals}]{
  \includegraphics[width=0.48\columnwidth]{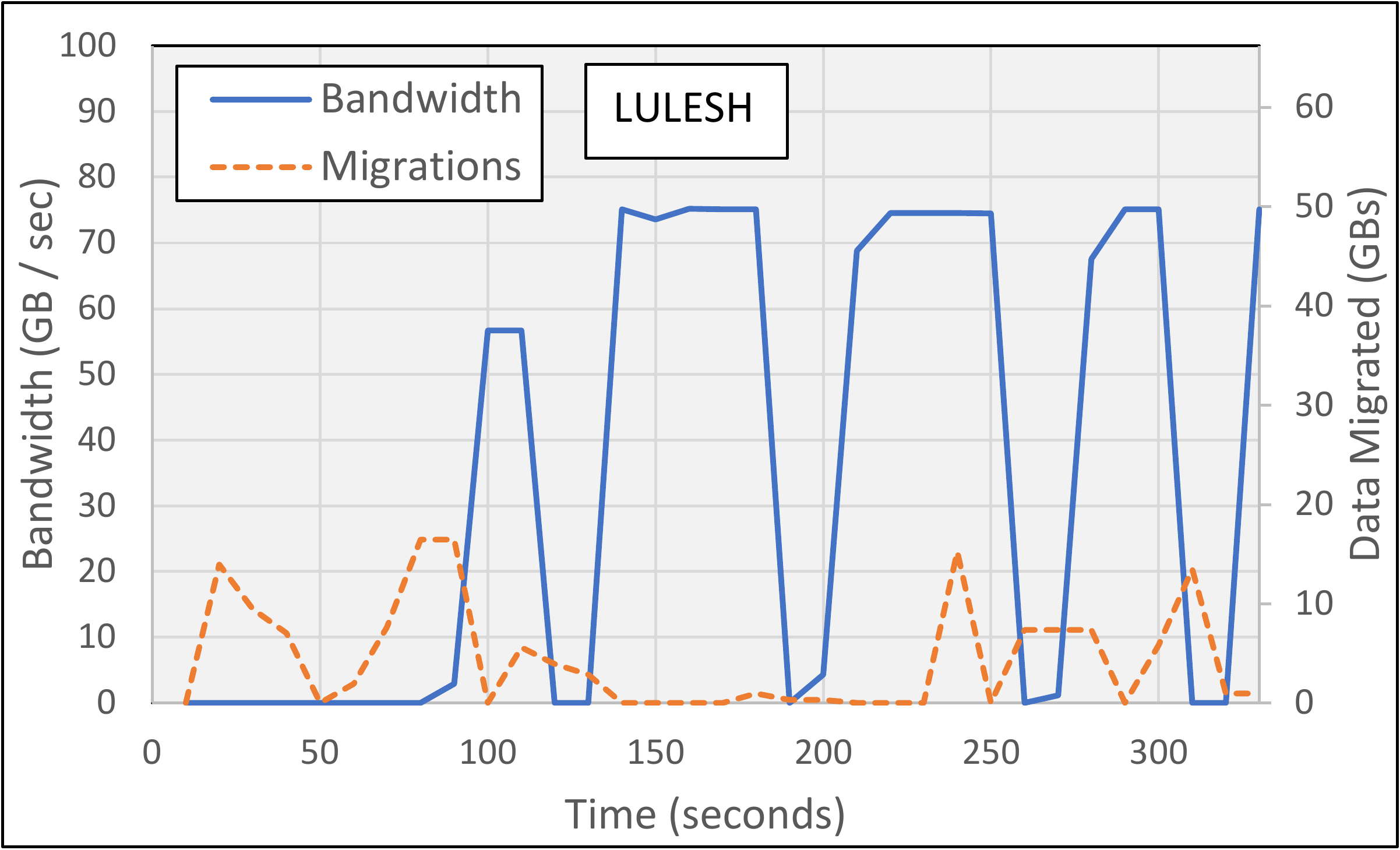}
}
\hfill%
\subfloat[AMG\label{fig:amg_bandwidth_intervals}]{
  \includegraphics[width=0.48\columnwidth]{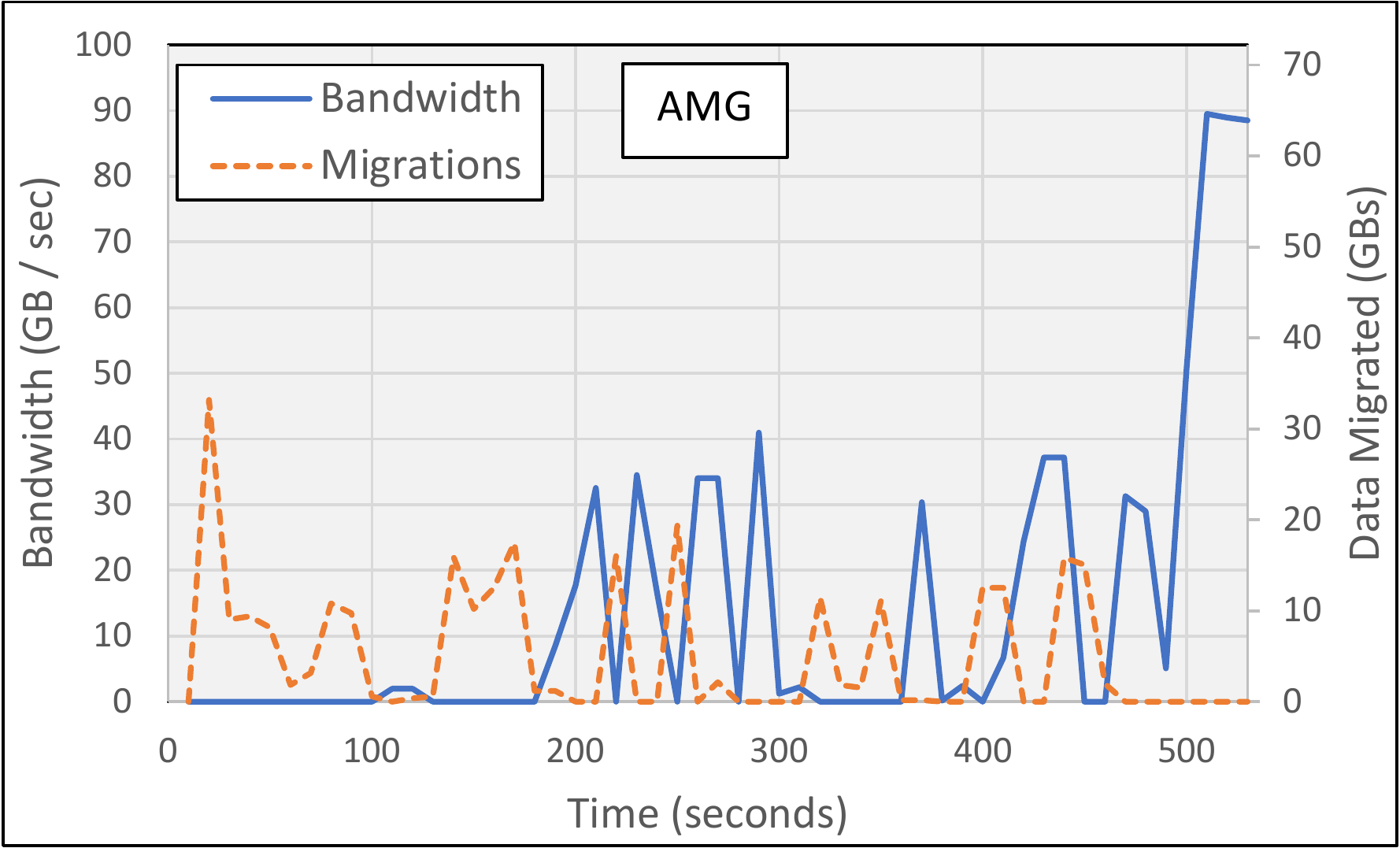}
}\\
\subfloat[SNAP\label{fig:snap_bandwidth_intervals}]{
  \includegraphics[width=0.48\columnwidth]{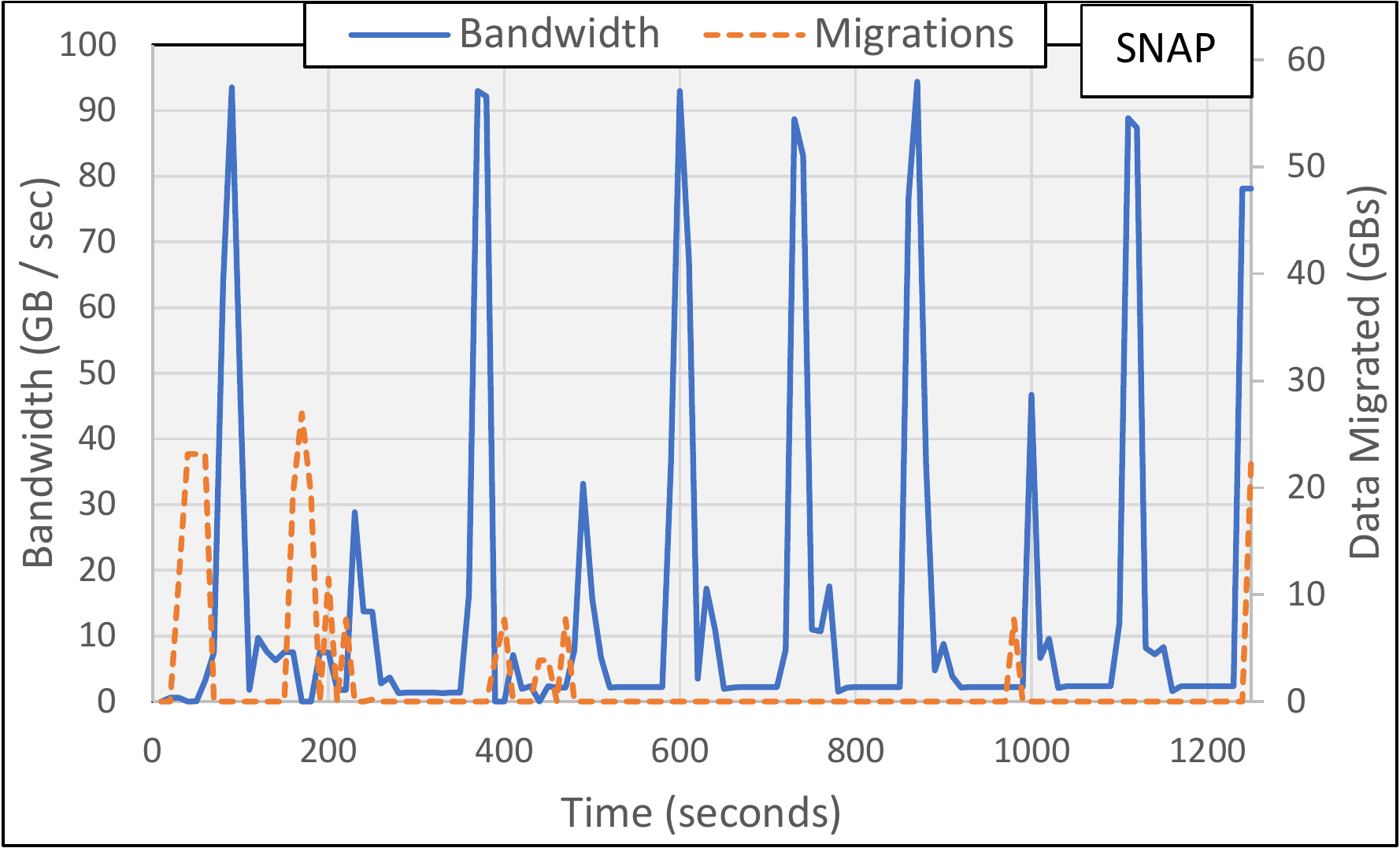}
}
\hfill%
\subfloat[QMCPACK\label{fig:qmcpack_bandwidth_intervals}]{
  \includegraphics[width=0.48\columnwidth]{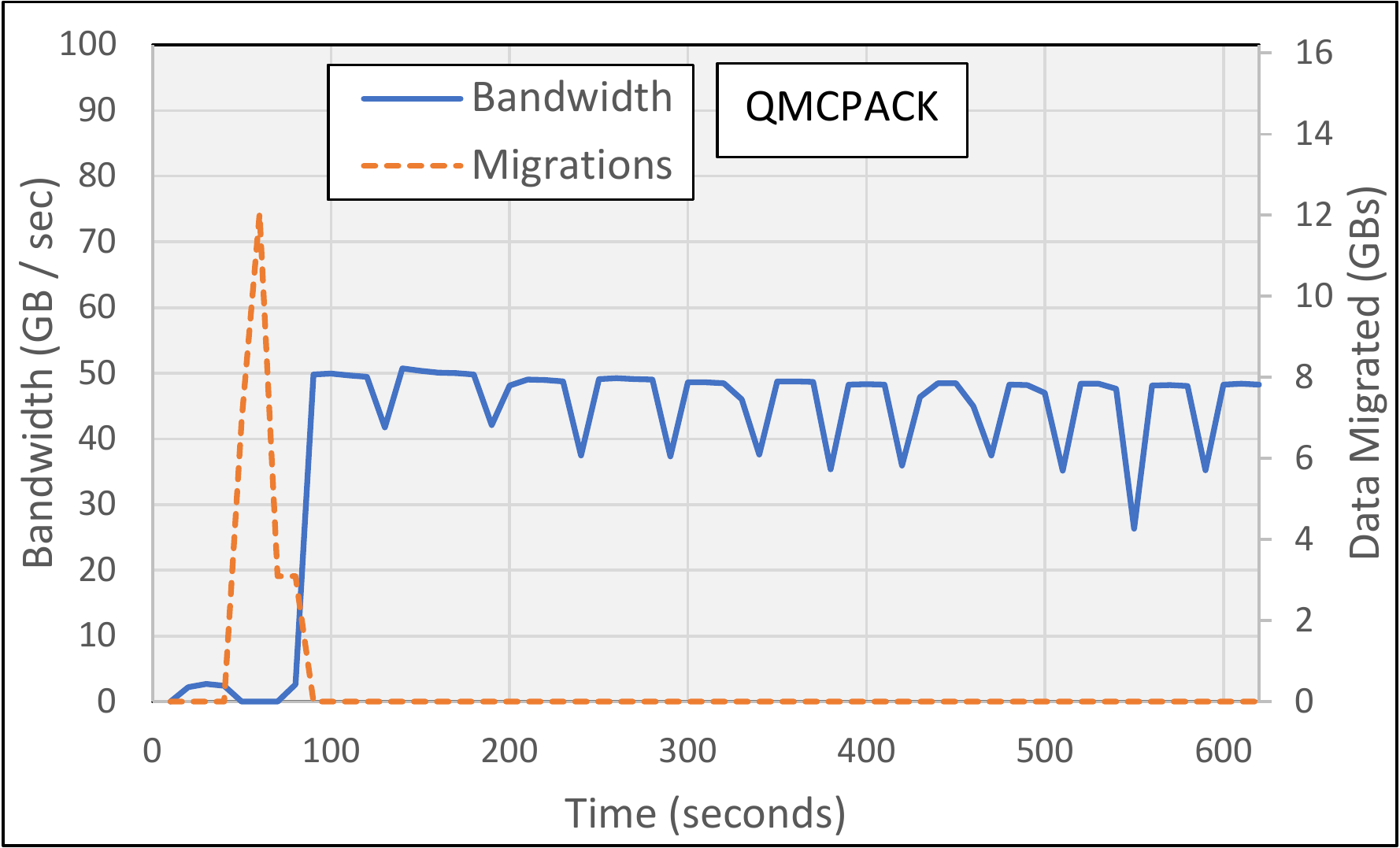}
}
\caption{Data bandwidth (GB / sec) and migrations (GBs) over time for four CORAL benchmarks with the medium input. The results are collected over 10s intervals during a run with the available DRAM limited to 50\% of the peak RSS of the application. Bandwidth is plotted on the left y-axis, which has a maximum value of 100 GB / sec, matching the maximum sustainable DRAM bandwidth on our platform. Data migrations are plotted on the right y-axis, which ranges from 0 to the peak RSS of the application.}
\label{fig:bandwidth_intervals}
\end{figure*}
\paragraph{Comparing the Offline and Online Approaches}
While there are a few cases where the online approach outperforms the offline approach, in general, the offline approach enables faster execution times than the online approach.
To explain why, we analyzed the individual online profiles of several of our selected benchmarks.
We found that the capacity and usage of data associated with each allocation site may shift substantially during early portions of the run, but after a short initial period, the sorted rank of each site and selection of sites assigned to the upper tier remains relatively stable from interval to interval.
Indeed, in every case we analyzed, the online approach converges to a solution similar to the offline approach after this short initial period.

To understand how this behavior impacts the operation of the online approach, consider Figure~\ref{fig:bandwidth_intervals}, which plots the total memory (DRAM + Optane\textsuperscript{TM}) bandwidth as well as the number of GBs migrated between memory tiers over time with the online approach for the four CORAL benchmarks with their medium input sizes.
Thus, we see that memory bandwidth for the online run is relatively low during early portions of the run, that is, until the runtime has enough profile information to make good data placement decisions.
Additionally, the majority of data migration occurs during this early period of relatively poor performance.
Even after the runtime identifies a good data placement strategy, later intervals may still sometimes interrupt the application to change data placement, but these migrations have relatively little impact on system bandwidth.

Hence, we find there are two main reasons that the performance of the online approach sometimes lags the offline approach for these benchmarks.
For one, the offline approach does not incur any overhead for profiling and is able to use an additional computing core, which would otherwise be used for profiling, for program execution.
We expect that future efforts could reduce this overhead by disabling some or all of the profiler after the initial startup period.
Additionally, even if the online approach converges to a similar or better solution than the offline approach, the application still executes with suboptimal data placement during the initial profiling period.
We found that this effect can have a considerable negative impact in some cases, as this initial period can be a significant portion of the total execution time for some benchmarks (e.g., AMG in Figure~\ref{fig:amg_bandwidth_intervals}).
Even with these limitations, the online approach is still preferable to the offline approach in cases where: 1) it is not feasible to collect or maintain offline profile data, or 2) it is not possible or very difficult to construct and profile program inputs that are representative of production execution.

\iffalse
\begin{itemize}
    \item The ability to run a larger variety of benchmarks (point out the inability
          for POP2 and QMCPACK to execute the old profiling techniques)
    \item A reduction in overhead with a geometric mean of 20.8\% for medium-sized workloads, 
          but only 1.76\% for larger ones.
    \item A dramatic reduction in profile interval time. This results in a geometric mean
          of 0.9 saved seconds for medium-sized workloads and 5.33 seconds for larger ones.
\end{itemize}
\fi

\vspace{-1ex}
\subsection{Performance Analysis with Large Memory Workloads}
\label{sec:large_eval}
We conclude our evaluation by examining the impact of guided data tiering on the CORAL benchmarks with the large and huge input sizes.
There are several benefits of evaluation with such large scale memory workloads.
Since these workloads require more memory capacity than there is available DRAM on our platform, there is no need to artificially limit the available capacity of the faster memory tier. 
As a result, data movement costs are also more realistic because the system may migrate data into and out of the entire DRAM tier.
Additionally, this approach allows for direct comparison between guided (and unguided) software-based data tiering approaches and the hardware-managed DRAM caching available on our platform.\footnote{In Intel\textsuperscript{\textregistered}'s literature, the hardware-managed caching option for our platform is referred to as \emph{memory mode}.}

\begin{wrapfigure}[14]{r}{0.5\textwidth}
  \vspace{-2ex}
  \centering
  \includegraphics[width=\linewidth]{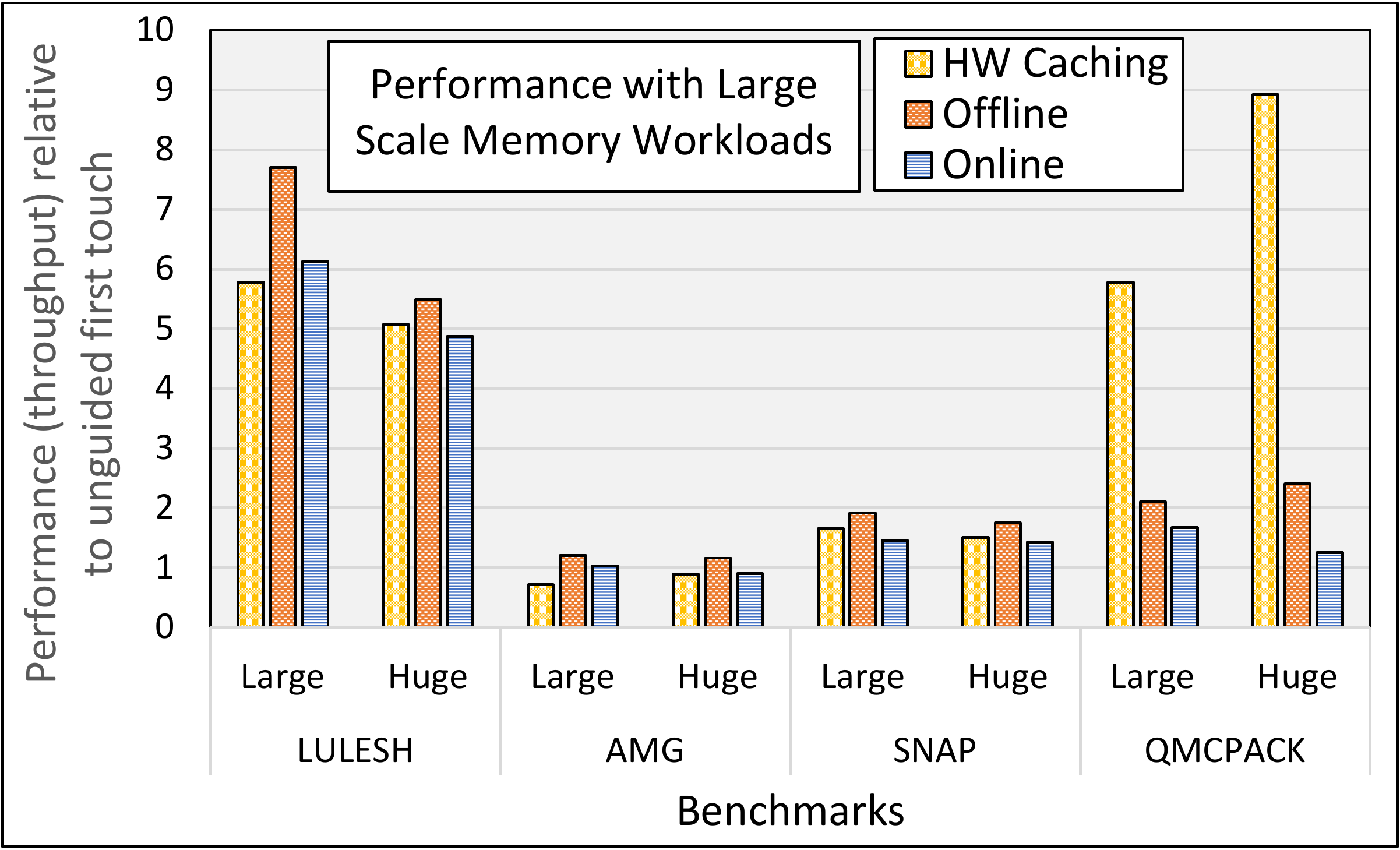}
  \captionof{figure}{Performance (throughput) of CORAL benchmarks with large and huge input sizes (higher is better).}
  \label{fig:large_huge_perf}
\end{wrapfigure}
Figure~\ref{fig:large_huge_perf} shows the performance of the four CORAL benchmarks with the offline and online approaches alongside the hardware-managed DRAM caching option on our platform.
Each bar shows throughput relative to the unguided first-touch configuration, and thus, higher is better.
We find that the offline and online approaches significantly outperform unguided first touch in almost every case.
In the best case, LULESH with the large input achieves speedups of more than 7.7x and 6.1x for the offline and online approaches, respectively.

For LULESH, AMG, and SNAP, the guided approaches achieve similar, or somewhat better, performance than the hardware-managed caching mode on our platform.
The biggest improvement comes with the large input of AMG, which is almost 70\% faster with the offline approach and 45\% faster with the online approach compared to HW caching.
For these cases, there is also significant potential for memory energy savings with the guided approaches, as hardware-managed caching typically generates much more data movement between tiers than software-based approaches.

For QMCPACK, however, hardware-managed caching is much more efficient than either guided approach, and achieves speedups ranging from 2.8x to 7x faster compared to the guided approaches.
This case highlights one of the remaining limitations of our profile guided approaches.
On further analysis, we found that the larger QMCPACK inputs use a single allocation site for the vast majority of their data allocations.
Specifically, this site creates between 60\% and 63\% of all resident program data during runs with the large and huge inputs.
Despite its large size, the data created at this site also exhibits the most frequent usage per byte relative to the other data in the application.
As a result, this data is almost always assigned to the DRAM tier during guided execution, even if a significant portion of it is relatively cold for some time.
Hence, while the guided approaches still outperform first touch, they lag the performance of the hardware-based approach, which is able to evict and replace cold data in the DRAM cache at a much finer granularity.

One approach we are considering to address this limitation is to fragment large sets of data created from the same site into separate sets based on different data features, such as the age of the data, or the PID of the allocating thread.
In this way, the runtime could distinguish data that share the same allocation context, while still limiting the number of data groups that it needs to profile and manage to guide data placement effectively.

\vspace{-1ex}
\section{Future Work}
We are currently pursuing several avenues of future work.
As described in Section~\ref{sec:large_eval}, we are experimenting with different options for clustering program data into groups with similar expected usage with the goal of enabling more effective prediction and management of memory usage.
In addition to tools that rely on source code analysis and compiler integration, we are also building versions of this approach that do not require access to program source code or recompilation.
These tools will leverage the application runtime and allocator to identify address ranges with similar expected usage, and will then send this information to a system-level daemon that tracks and manages data allocations and placement for multiple processes.
Using these tools, we also plan to conduct experiments to examine and quantify the benefits of high-level knowledge of application source code for guiding data placement.

At the same time, we are experimenting with features of managed language runtime systems, such as the Java virtual machine (JVM), to further enhance guided data management.
High-level language VMs (HLL-VMs) offer a number of features that can simplify (and often boost the efficiency of) classifying and migrating heap data.
Specifically, these systems typically shield applications from directly accessing the locations of objects on the heap, thereby freeing them from the need to update references with relocated addresses when objects migrate.
Other features of HLL-VMs, such as garbage collection, and emulation engines that are designed for FDOs, can also make guidance-based data management easier to deploy and more effective for managed language applications.

This study targeted a state-of-the-art heterogeneous memory platform with conventional DDR4 SDRAM and non-volatile Optane\textsuperscript{TM} RAM.
As we take this work forward, we will modify our tools and framework for use with other architectures and emerging technologies, including systems with three or more tiers of distinct memory hardware, and will explore the challenges and opportunities that arise from guiding data management on more complex memory architectures.

\vspace{-1ex}
\section{Conclusions}
This work develops the first ever fully automatic and online profile guided data tiering solution for heterogeneous memory systems.
It extends our previous \emph{offline} profiling-based approach with new techniques to collect data tiering guidance with very low, and often negligible, performance overhead.
It also develops a novel online algorithm that periodically analyzes this high-level information and uses it to steer data allocation and placement across a heterogeneous memory architecture.
The evaluation shows that this approach significantly outperforms unguided data placement on a state-of-the-art Intel\textsuperscript{\textregistered} platform with DDR4 SDRAM and Optane\textsuperscript{TM} NVRAM, with speedups ranging from $1.4x$ to $7x$ for a standard set of HPC workloads.
Additionally, we find that, aside from a short startup period needed for convergence, the online approach achieves performance similar to that of a well-tuned offline approach.
However, because it adapts automatically to the program as it runs, it does not need to collect or store profile information from a separate execution, which can be unwieldy, and may lead to stale or unrepresentative profile guidance.

%%%%%%%%%%%%%%%%%%%%%%%%%%%%%%%%%%%%%%%%%%%%%%%%%%%%%%%%%%%%
% ACKNOWLEDGEMENTS
%%%%%%%%%%%%%%%%%%%%%%%%%%%%%%%%%%%%%%%%%%%%%%%%%%%%%%%%%%%%
%\begin{acks}
%\end{acks}

%%%%%%%%%%%%%%%%%%%%%%%%%%%%%%%%%%%%%%%%%%%%%%%%%%%%%%%%%%%%
% BIBLIOGRAPHY
%%%%%%%%%%%%%%%%%%%%%%%%%%%%%%%%%%%%%%%%%%%%%%%%%%%%%%%%%%%%
\bibliographystyle{ACM-Reference-Format}
\bibliography{bibliography/introduction,bibliography/related_work,bibliography/offline,bibliography/online,bibliography/experimental,bibliography/evaluation}

\end{document}